\documentclass[12pt]{article}

\usepackage[dvips]{graphicx}

\newcommand{\lsim}{\raisebox{-4pt}{$\,\stackrel{\textstyle
                                                         <}{\sim}\,$}}

\newcommand{\nn}{\nonumber}   
\newcommand{\be}{\begin{equation}}        
\newcommand{\ee}{\end{equation}}     
\newcommand{\ba}{\begin{eqnarray}}  
\newcommand{\ea}{\end{eqnarray}}     
\newcommand{\req}[1]{(\ref{#1})}
\def\={\,=\,}                      
\newcommand{\ci}[1]{\cite{#1}}     

\def\mev{~{\rm MeV}}
\def\gev{~{\rm GeV}}     
\def\kev{~{\rm keV}}
\def\ale{\alpha_{\rm elm}}
\def\als{\alpha_{\rm s}}

\def\LQCD{\Lambda_{\rm QCD}}  
\def\muF{\mu_F}             
\def\muR{\mu_R}       
\def\muO{\mu_{0}}             
\def\xb{\bar{x}}
\newcommand{\tw}{\textwidth}   
\def\vk{{\bf k}_{\perp}}     
\def\vbs{{\bf b}}               
\newcommand{\da}{{distribution amplitude}}  
\newcommand{\wf}{wave function}              
\newcommand{\ov}[1]{\overline#1}
\def\qh{\hat{q}}    
\def\bh{\hat{b}}

\begin{document} 

\thispagestyle{empty}
\begin{flushright}
WU B 10-36
\end{flushright}
\vskip 10mm

\begin{center}
{\Large\bf The form factors for the photon to pseudoscalar meson
  transitions - an update}
\vskip 10mm

P.\ Kroll \\[1em]
{\small \it Fachbereich Physik, Universit\"at Wuppertal, 42097 Wuppertal,
  Germany \\
and\\
Institut f\"ur Theoretische Physik, Universit\"at Regensburg,\\
          93040 Regensburg, Germany}\\
\end{center}
\vskip 5mm 
\begin{abstract}
 The form factors for the transitions $\pi\gamma$, $\eta\gamma$,
$\eta^\prime\gamma$ and $\eta_c\gamma$ are analyzed within the modified
perturbative approach in which quark transverse degrees of freedom are
retained. The results for the form factors are compared to experiment in
detail. As compared to previous calculations within the same approach 
only little modifications of the meson \da s are required in general in 
order to achieve agreement with experiment. Only for the $\pi\gamma$ form
factor a strong contribution from the second Gegenbauer term is found.
It also commented on the case of two virtual photons and on the transition 
form factors in the time-like region.
\end{abstract}
  
\centerline{December, 12  2010} 

\section{Introduction}
\label{sec:introduction}
The simplest exclusive observable is the $\pi\gamma$ transition form factor, 
$F_{\pi\gamma}$. It has been shown \ci{lep79} that its behavior for large 
photon virtuality, $Q^2$, is determined by the operator product expansion of 
the product of two electromagnetic currents near the light cone. The only 
soft physics information required in the calculation of the form factor is \
the pion wave function. It is however theoretically not understood what large 
$Q^2$ means or, in other words, at which value of $Q^2$ this collinear 
factorization approach can be applied. This is still to be decided by 
comparison with experiment.

In 1995 the CLEO collaboration \ci{savinov} showed the first data on the 
$\pi\gamma$ transition form factor at fairly large values of $Q^2$, actually 
up to $8\,\gev^2$. These data provoked an enormous theoretical activity. It 
seemed that collinear QCD provides a large contribution to the transition
form factor. The data lie only about $20\%$ below the asymptotic limit of 
that form factor, namely $Q^2F_{\pi\gamma}\to \sqrt{2}f_\pi$ where 
$f_\pi$ ($=131\,\mev$) is the pion's decay constant. Next-to-leading order 
(NLO) corrections \ci{NLO} account for about a half of the difference between
the asymptotic results and the data. For the remaining difference several 
explanations were proposed: a pion \da{} slightly different from the 
asymptotic one ($\Phi_{AS}=6x(1-x)$), low renormalization scales which 
enhance the NLO corrections, power corrections or quark transverse momenta 
to mention a few mechanisms. Since all these mechanisms have to provide only 
little effects the $\pi\gamma$ transition form factor was believed to be the 
theoretically best understood exclusive observable.
 
Recently, this believe has been ruined. The BaBar collaboration \ci{babar09}
has measured the $\pi\gamma$ transition form factors up to about
$35\,\gev^2$. While these data agree with the CLEO data \ci{savinov,CLEO} 
below $8\,\gev^2$ they reveal an expectedly large rise towards large $Q^2$. 
This behavior is in dramatic conflict with dimensional scaling and turned 
previous calculations obsolete. As the CLEO data in 1995 the BaBar data 
renewed the interest in this quantity and many paper papers have already 
been devoted to its theoretical analysis, e.g.\ \ci{rad09} - \ci{teryaev}.  

Here, in this article, it is proposed to employ the modified perturbative 
approach (MPA) invented by Sterman and collaborators \ci{botts89,li92}. 
This approach which bases on  $\vk$ factorization, has been used before in 
the analysis of the CLEO data \ci{JKR95,raulfs95}. It will be shown below 
that a good fit to the BaBar data can be achieved with this approach, only 
a few higher order terms in the Gegenbauer representation of the \da s have 
to be taken into account now. Arguments will also be given why in the $Q^2$ 
range covered by the CLEO data the simple asymptotic \da{} suffices for a 
fair fit to the data. In the next section the MPA will be described in some 
detail and its properties discussed. Next, in Sect.\ \ref{sec:babar}, the 
actual analysis of the data on the $\pi\gamma$ form factor will be presented 
and the results compared with other theoretical approaches. Sect.\
\ref{sec:eta} is devoted to the analysis of the $\eta\gamma$ and 
$\eta^\prime\gamma$ transition form factors which have been measured by CLEO 
\ci{CLEO} and the L3 collaboration \ci{L3} previously and by BaBar 
\ci{BaBar-prel} recently. The Babar data \ci{babar10} on the $\eta_c\gamma$ 
transition form factor will be analyzed in Sect.\ \ref{sec:etac} briefly. 
Comments on the case of two virtual photons and on the form factors in the 
time-like region will be presented in the respective Sects.\ \ref{sec:two} 
and \ref{sec:timelike} before the papers ends with a summary (Sect.\ 
\ref{sec:summary}). Details of the Sudakov factor will be presented in an 
Appendix.

\section{The modified perturbative approach}
\label{sec:mpa}
An alternative to the usual collinear factorization is the transverse-momentum
($\vk$)-factorization. For hard exclusive processes this type of factorization
has been proposed by Sterman and collaborators \ci{botts89,li92}. Further 
arguments for the validity of $\vk$-factorization are given in \ci{nagash} for 
the case of the transition form factors. Nevertheless a rigorous proof of 
$\vk$-factorization does not exist as yet. The basic idea of
$\vk$-factorization is to retain the quark transverse degrees of freedom in
the hard scattering. This however implies that quarks and antiquarks are
pulled apart in the transverse-configuration or impact-parameter space. The 
separation of color sources is accompanied by the radiation of gluons. Based
on previous work by Collins {\it et al} \ci{soper81,soper82,sterman85} the 
corrections to the hard scattering process due to gluon radiation have been 
calculated in Ref.\ \ci{botts89} in axial gauge using resummation techniques 
and having recourse to the renormalization group. These radiative corrections 
comprising re-summed leading and next-to-leading logarithms which are not
taken into account by the usual QCD evolution, are presented in the form of a 
Sudakov factor in the impact-parameter plane.
  
The $\vk$-factorization combined with a Sudakov factor is termed the MPA. Its
advantage is that the end-point regions where one of the parton momentum
fractions tends to zero, are strongly damped. Large contributions to the form
factor accumulated in the soft end-point regions would render the use of 
perturbation theory inconsistent. As has been pointed out by Isgur and
Llewellyn-Smith~\ci{IL89} this is frequently the case in collinear factorization
at experimentally accessible values of momentum transfer, typically a few 
$\gev^2$. Another advantage of the MPA is that the renormalization scale,
$\muR$, can be chosen to be momentum-fraction dependent in order to eliminate 
large logarithms from higher-order perturbative corrections. Eventual 
$\als$-singularities are canceled by the Sudakov factor without introducing 
additional {\it ad hoc} cut-off parameters as for instance a parton mass. 
Thus, the MPA provides well-defined expressions for form factors and
amplitudes of hard scattering processes: the perturbative contributions can be
calculated in a self-consistent way, even for momentum transfers as low as a 
few $\gev^2$.

\begin{figure}
\begin{center}
\includegraphics[width=.65\textwidth,bb=135 550 578
  653,clip=true]{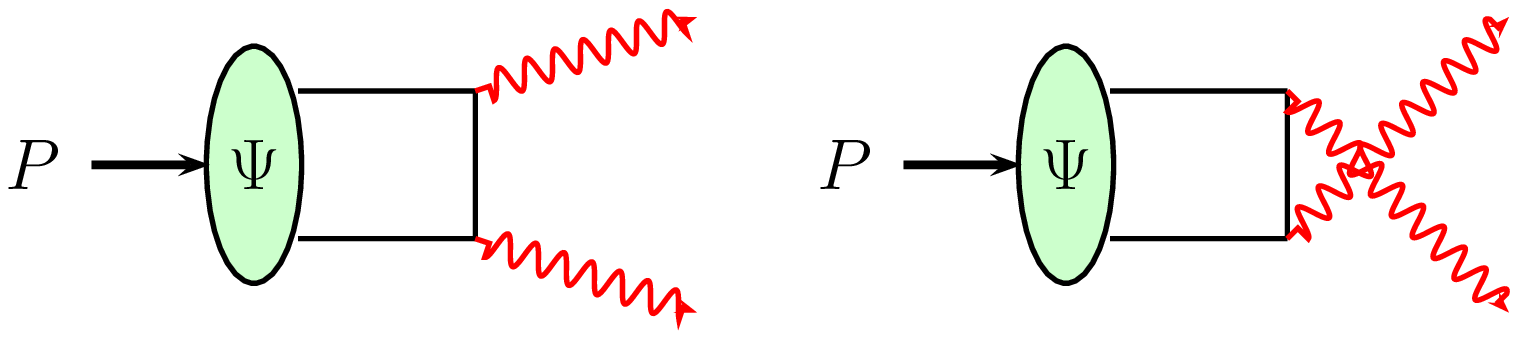}
\end{center}
\caption{Lowest order Feynman graphs for the $P\gamma$ transition form
  factor.}
\label{fig:LO-graphs}
\end{figure}

According to \ci{JKR95,raulfs95} the form factor for a transition from a 
photon to a pseudoscalar meson (P) reads
\be
F_{P\gamma}(Q^2) \= \int dx \frac{d^2\vbs}{4\pi}\,
\hat{\Psi}_P(x,-\vbs,\muF) \hat{T}_H^P(x,\vbs,Q,\muR) e^{-S(x, b, Q, \muR, \muF)}\,,
\label{eq:FF-mpa}
\ee
within the MPA. Since the Sudakov exponent $S$ is given in the 
impact-parameter ($\vbs$) space, see App. A, it is convenient to work in that
space which is canonically conjugated to the $\vk$-space. In the convolution 
formula \req{eq:FF-mpa} $\hat{T}_H$ is the Fourier transform of the 
momentum-space hard scattering amplitude $T_H$ evaluated to lowest order 
perturbative QCD from the Feynman graphs displayed in Fig.\
\ref{fig:LO-graphs} but taking into account quark transverse momenta
\be
T_H^P\=\frac{4\sqrt{6}\, C_P}{xQ^2+\vk^2}\,,
\label{eq:hard-amplitude}
\ee
Here $C_P$ is a charge factor. For instance, for the pion it reads 
$C_\pi=(e_u^2-e_d^2)/\sqrt{2}$ where $e_a$ denotes the charge of a flavor-$a$
quark in units of the positron charge. The Fourier transform of
\req{eq:hard-amplitude} is 
\be
\hat{T}_H^P \= \frac{2\sqrt{6}\,C_P}{\pi}\,K_0(\sqrt{x}Qb)\,,
\label{eq:th}
\ee
where the Fourier transform is defined by 
\be
\hat{f}(\vbs)\= \frac1{(2\pi)^2}\, \int d^2\vk \exp{[-i\, \vbs \cdot \vk]}
f(\vk)\,.
\ee
The function $K_0$ denotes the modified Bessel function of order zero. 

Another item in \req{eq:FF-mpa} is $\hat{\Psi}_P$, the light-cone wave
function of the meson P in the impact-parameter space. In the original version 
of the MPA \ci{li92} this \wf{} is assumed to be just the meson \da{} $\Phi_P$. 
As argued in \ci{li92} the Sudakov factor $e^{-S}$ can be viewed as the 
perturbatively generated transverse part of the \wf{}. For low and intermediate 
values of $Q^2$, however, the non-perturbative intrinsic $\vbs$- or 
$\vk$-dependence of the light-cone \wf{} cannot be ignored as has been pointed 
out in \ci{jakob93}. The inclusion of the transverse size of the meson extends 
considerably the region in which the perturbative contribution to the form
factor can be calculated. As in \ci{raulfs95,jakob93} the \wf{} is
parameterized in the form
\be
\hat{\Psi}_P(x,\vbs,\mu_F) \= 2\pi
\frac{f_P}{\sqrt{6}}\,\Phi_P(x,\muF)\,
                            \exp{\Big[-\frac{x\xb b^2}{4 \sigma_P^2}\Big]}\,.
\label{eq:Gaussian}
\ee
The \da{}, $\Phi_P$, possesses a Gegenbauer expansion
\be
\Phi_P(x,\muF) \= \Phi_{\rm AS} \left[1 + \sum_{n=2,4,\cdots} a_n(\muO) 
                \Big(\frac{\als(\muF)}{\als(\muO)}\Big)^{\gamma_n}
                C_n^{3/2}(2x-1)\right]\,,
\label{eq:distribution-amplitude}
\ee
where the evolution of the expansion parameters $a_n$ from an initial scale,
$\muO$, to the factorization scale, $\muF$, is controlled by
the anomalous dimensions ($n\geq 0$)
\be
\gamma_{n+2} \= \gamma_n + 4 \frac{C_F}{\beta_0}
\frac{(2n+5)(n^2+5n+5)}{(n+1)(n+2)(n+3)(n+4)}\,.
\label{eq:anomalous-dimensions}
\ee
Here, $\gamma_0=0$, $C_F=4/3$ and $\beta_0=11-2/3n_f$, $n_f$ is the number of
active flavors. For the cases of interest in this work the \da{} is symmetric 
under the replacement of the momentum fraction $x$ by $\xb=1-x$. This symmetry
is already taken into account in \req{eq:hard-amplitude}. The transverse size 
parameter $\sigma_P$ being related to the r.m.s transverse momentum by (with
$a_0=1$)
\ba
\langle \vk^2 \rangle &=& \frac{6\pi^2 f_P^2}{P_{q\bar{q}}} \int dx x^2\xb^2
\sum_{n=0,2,\cdots} \Big[a_n^2 (C_n^{3/2})^2 + 2 a_n a_{n+2}C_n^{3/2}
C_{n+2}^{3/2}\Big]                      \nn\\
   &=& \frac{\pi^2 f_P^2}{5 P_{q\bar{q}}}\,\Big[1- \frac67 a_2(1-2a_2)
              -\frac{150}{77} a_4(a_2-\frac{35}{26} a_4) + \ldots \Big]  
\label{eq:rms-k}
\ea
For later use the probability of the meson's valence Fock state is also quoted: 
\be
P_{q\bar{q}} \= \int dx \,\frac{d^2b}{4\pi} \mid \hat{\Psi}_P(x,\vbs)\mid^2 \=
3\pi^2 f_P^2 \sigma_P^2\, \sum_{n=0,2,\cdots} a_n^2 \,\frac{(n+1)(n+2)}{2n+3}\,.
\label{eq:probability}
\ee
The requirement $P_{q\bar{q}}\leq 1$ leads to bounds on the Gegenbauer
coefficients or strictly speaking, on the products $a_n\sigma_P$
\be
\mid a_n\sigma_P\mid \leq \frac1{\pi f_P}\,
\sqrt{\frac{2n+3}{3(n+1)(n+2)}}\,.
\label{eq:bound}
\ee 
For instance, if $\sigma_P=0.5\, (1.0)\,\gev^{-1}$ at the initial scale
$\mu_0$ one finds
\be
a_2(\mu_0)\leq 2.14\, (1.07)\,, \qquad a_4(\mu_0)\leq 1.70\, (0.85)\,.
\ee

The Sudakov exponent $S$ in \req{eq:FF-mpa} which comprises the characteristic
double logarithms produced by overlapping collinear and soft divergencies for 
massless quarks, is given in App.\ A. The impact parameter $\vbs$ which
represents the transverse separation of quark and antiquark, acts as an 
infrared (IR) cut-off parameter~\footnote{
   For a more complicated system like the proton's electromagnetic form factor,
   there are several $\vbs$'s. In order to cancel the $\als$ singularities the
   $\vbs$'s have to be chosen appropriately \ci{bolz95a}.}
\ci{soper81,soper82,sterman85}. Thus, $1/b$ in the Sudakov exponent marks
the interface between the non-perturbative soft momenta which are
implicitly accounted for in the hadron \wf, and the contributions from soft 
gluons, incorporated in a perturbative way in the Sudakov factor. Obviously, 
the IR cut-off serves at the same time as the gliding factorization scale
\be
 \muF \= 1/b
\ee
to be used in the evolution of the \wf. In accord with this interpretation the
entire Sudakov factor is continued to zero whenever $b > 1\LQCD$. In this 
large-$b$ region the wavelength of the radiative gluon is larger than
$1/\LQCD$. Because of the color neutrality of the hadron such gluons cannot 
resolve the hadron's quark distribution; hence radiation is damped. Soft 
gluons with wavelength larger than $1/\LQCD$ are therefore to be excluded from 
perturbation theory; they have to be absorbed into the soft \wf. 

\begin{figure}
\begin{center}
\includegraphics[width=.65\textwidth,bb=113 310 494 580,clip=true]
{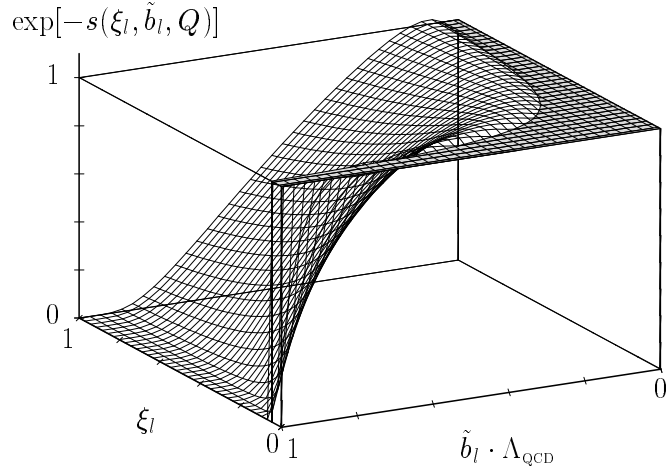}
\end{center}
\caption{The exponential of the Sudakov function $s(\xi_l,\tilde{b}_l,Q)$ vs
  $\xi_l=\xi$ and $\tilde{b}=b$ for $Q=30\LQCD$. The hatched area
  indicates the hard scattering region. The plot is taken from \ci{bolz95a}.}
\label{fig:sudakov}
\end{figure}

Radiative corrections with wavelengths between the IR cut-off and the upper 
limit $\sqrt{2}/\xi Q$ yield to suppression through the Sudakov factor as is
evident from the integration limits in \req{eq:s-integral} ($\xi$ is either
$x$ or $1-x$). Gluons with even shorter wavelengths are regarded as hard ones 
which are considered as higher-order perturbative corrections of the hard 
scattering and, hence, are not part of the Sudakov factor. For that reason, 
the Sudakov function $s(\xi,b,Q)$ which is defined in \req{eq:s} and is 
related to $S$ according to \req{eq:S}, is set equal to zero \ci{li92}  
whenever 
\be
\xi\leq \frac{\sqrt{2}}{Qb}\,.
\label{eq:requirement}
\ee  
In Fig.~\ref{fig:sudakov} the exponential of the Sudakov function 
$\exp[-s(\xi,b,Q)]$ for $Q=30\: \Lambda _{\rm{QCD}}$ is displayed.
The properties of the Sudakov function lead to an asymptotic damping of any
contribution except those from configurations with small quark-antiquark 
separations, i.e.\ for $\ln{Q^2}\to\infty$ the limiting behavior of the
transition from factors in collinear factorization emerges, for instance,
$F_{\pi\gamma}\to \sqrt{2}f_\pi /Q^2$.  

In a NLO calculation of the $P\gamma$ form factor or, say, in the case of
the electromagnetic pion form factor a momentum-fraction dependent
renormalization scale leads to a singular $\als$ in the limit $\muR\to\LQCD$. 
This singularity is canceled in the MPA: whenever $\als$ tends to infinity the 
Sudakov factor ${\rm e}^{-S}$  rapidly decreases to zero. As one may suspect
from Fig.~\ref{fig:sudakov} this does not seem to be the case in the region 
determined by $b\LQCD \to 1$ and simultaneously $\xi\leq \sqrt{2}\LQCD/Q$,
where $\exp[-s(\xi,b,Q)]$ is fixed to unity. However, in this region the other
Sudakov function in \req{eq:S}, namely $s(1-\xi,b,Q)$, provides the required
suppression. The logarithmic singularities $[\ln{(1/(b\LQCD)^2}]^{-\gamma_n}$
which arise from the evolution of the \wf{} and which become worse with
increasing Gegenbauer index are analogously cancelled by the Sudakov factor.

In analogy to the case of the pion's electromagnetic form factor \ci{li92} 
the maximum of the longitudinal scale appearing in the scattering amplitude 
\req{eq:hard-amplitude} and the transverse scale
\be
\muR \= \max(\sqrt{x}Q,1/b)\,,
\label{eq:renormalization}
\ee 
is chosen as the renormalization scale \ci{JKR95,raulfs95}. Although to lowest
order there is no $\als$ in the hard scattering amplitude for the $P\gamma$ 
transition form factor, it nevertheless depends on $\mu_R$. Indeed, as 
discussed above, the Sudakov factor comprises the gluonic radiative corrections 
for scales between $1/b$ and $\xi Q/\sqrt{2}$. Hence, the latter scale
specifies the onset of the hard scattering regime.  

\begin{figure}[th]
\begin{center}
  \includegraphics[width=.65\tw, bb=114 369 546 713,clip=true]
                                              {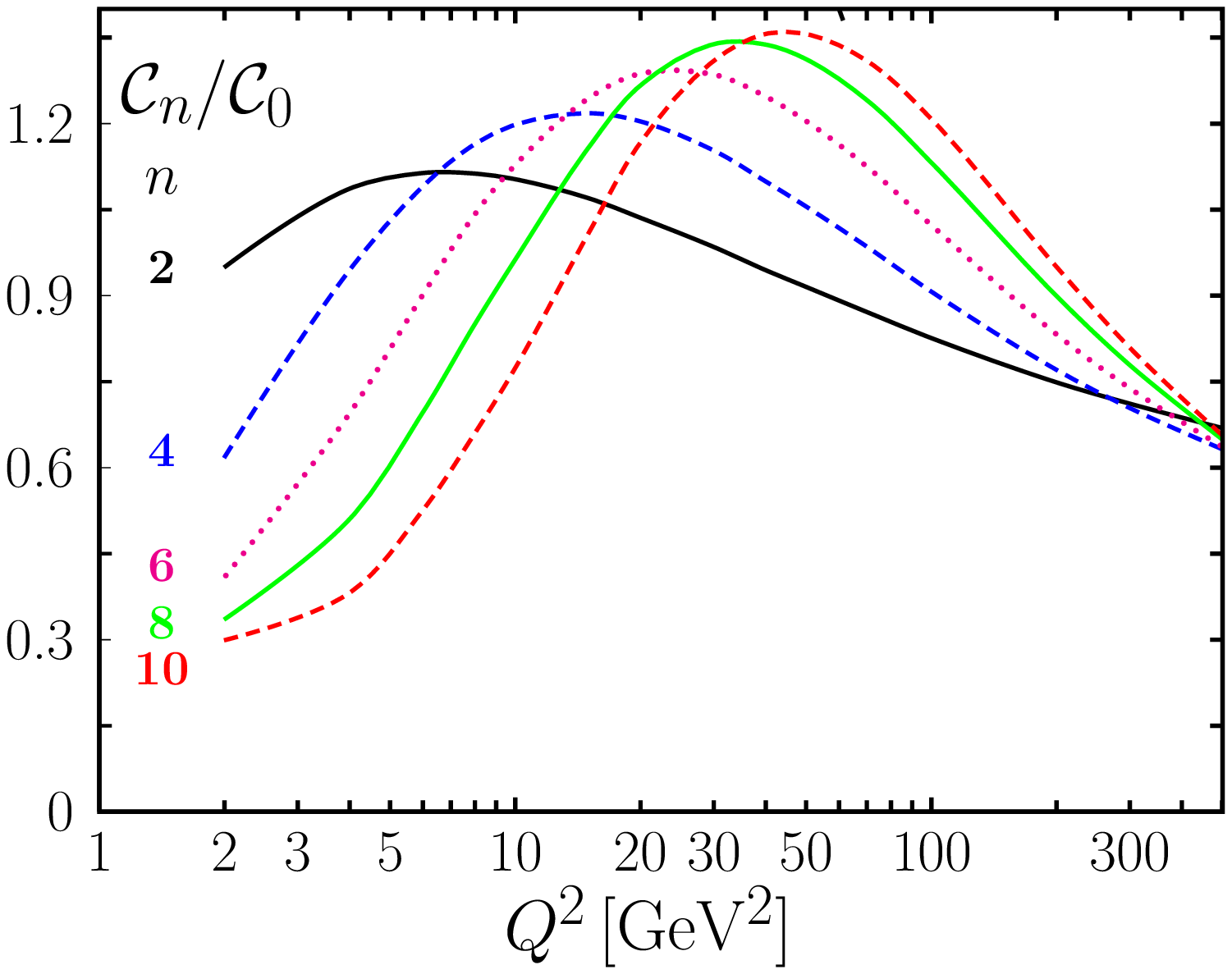} 
\end{center}
\caption{The relative strength of the contributions from the $n$-th Gegenbauer
 term to the $P\gamma$ form factor, scaled by the $n=0$ term, at the scale 
  $1.0\,\gev$. Intrinsic $\vk$ is omitted. (Colors online)}
\label{fig:gegenbauer}
\end{figure}

Inserting the Gegenbauer decomposition \req{eq:distribution-amplitude} of the
meson \da{} into \req{eq:FF-mpa} and integrating term by term one can write 
the $P\gamma$ transition form factor as
\be
Q^2 F_{P\gamma}\= 6 C_P f_P \,
\sum_{n=0,2,4\ldots} a_n(\muO)\, {\cal C}_n(Q^2,\muO,\sigma_P)\,
\label{eq:Cn}
\ee
The functions ${\cal C}_n$ incorporate the integrals over the product of the
$n$-th Gegenbauer component of the meson \wf{}, the hard scattering amplitude 
and the Sudakov factor as well as the change of the Gegenbauer coefficients 
with the factorization scale. The evolution is worked out with the 1-loop
expression for $\als$ using $\LQCD=0.181\,\gev$ and four flavors. 
This value of $\LQCD$ is used throughout the paper except stated otherwise. 
In \req{eq:Cn}  $\muO$ merely acts as the scale at which the Gegenbauer
coefficients of the \da{} are quoted; the transition form factor is
independent of it. Numerical results for the ratios ${\cal C}_n/{\cal C}_0$ 
obtained under neglect of the intrinsic transverse momentum (i.e.\ for 
$\sigma_P\to\infty$), are shown in Fig.\ \ref{fig:gegenbauer}. A remarkable 
property of the MPA is to be observed from this figure: For $n>0$ and low 
$Q^2$ the Gegenbauer terms ${\cal C}_n$ are suppressed as compared to the 
lowest one, ${\cal C}_0$. The strength of the suppression grows with the
Gegenbauer index. A closer inspection of the functions ${\cal C}_n$ in 
\req{eq:Cn} reveals that the Sudakov factor in conjunction with the hard 
scattering amplitude provides a series of power suppressed terms which come 
from the region of soft quark momenta ($x, 1-x \to 0$) and grow with the 
Gegenbauer index \ci{BDK}. With increasing $Q^2$ the higher Gegenbauer terms 
become gradually more important. At very large $Q^2$ however the evolution 
of the expansion coefficients again suppresses all ${\cal C}_n$ except 
${\cal C}_0$. Asymptotically, one has ${\cal C}_0=1$ and ${\cal C}_n=0$ for 
$n>0$ and the asymptotic limit of the transition form factor emerges. The 
behavior of the ${\cal C}_n$ in the MPA is very different from that in the 
collinear factorization approach. To LO for instance one has 
\be
{\cal C}_0^{\rm coll}\=1\,, \qquad 
{\cal C}_n^{\rm coll}\=[\als(\mu_F)/\als(\mu_0)]^{\gamma_n}\,. 
\ee

The intrinsic transverse momentum dependence of the \wf, on the other hand, 
also provides a series of power suppressed terms but these come from all 
$x$ and do not grow with $n$ \ci{BDK}. These power corrections superimposes
those generated by the Sudakov factor in combination with the hard scattering 
amplitude. For small values of the parameter $\sigma_P$ the power corrections 
from the intrinsic transverse momentum dependence are rather strong and
weaken the relative supppression of the higher Gegenbauer terms, in 
particular for small $n$. In exchange for this the ${\cal C}_0$ term itself 
increases more sharply with rising $Q^2$ towards the asymptotic limit 
${\cal C}_0=1$.

This property of the MPA explains why the CLEO data on the $\pi\gamma$
transition form factor \ci{CLEO} are well described by the asymptotic 
\da{} as shown in \ci{raulfs95} (with $\sigma_\pi=0.861\,\gev^{-1}$,
$\LQCD=0.20\,\mev$), see Fig.\ \ref{fig:fit-a4}. With the BaBar 
data \ci{babar09} at disposal which extend to much larger values of 
$Q^2$ and do exceed the asymptotic limit $\sqrt{2} f_\pi$ , higher 
Gegenbauer terms can no more be ignored; they are now required for a 
successful description of the transition form factors. What can be 
learned about the higher Gegenbauer terms from the BaBar data will be 
discussed in the next sections.

\section{Confronting with the BaBar data on $\pi\gamma$ 
transitions}
\label{sec:babar}
Now, having specified all details of the MPA, one can analyze the $\pi\gamma$
form factor by inserting \req{eq:Gaussian} and \req{eq:distribution-amplitude} 
into \req{eq:FF-mpa} and fitting the Gegenbauer coefficients to the CLEO
\ci{CLEO} and BaBar \ci{babar09} data. From a detailed examination
of the data it becomes apparent that besides the transverse size parameter
only one Gegenbauer coefficient can safely be determined. If more coefficients 
are freed the fits become unstable. The coefficients acquire unphysically
large absolute values between 1 and 10 (often in conflict with the bound 
\req{eq:bound}) and with alternating signs leading to strong compensations 
among the various terms. The reason for this fact is that the data show the 
tendency of a sharp increase of the slope in the vicinity of 
$Q^2\simeq 10\,\gev^2$ while theory for $a_n>0$ provides a slowly decreasing
slope. With large absolute values and alternating signs of the coefficients 
a slightly increasing slope is produced for a finite range of $Q^2$. 

\begin{figure}[th]
\begin{center}
\includegraphics[width=.60\tw, bb=105 385 569 721,clip=true]
                {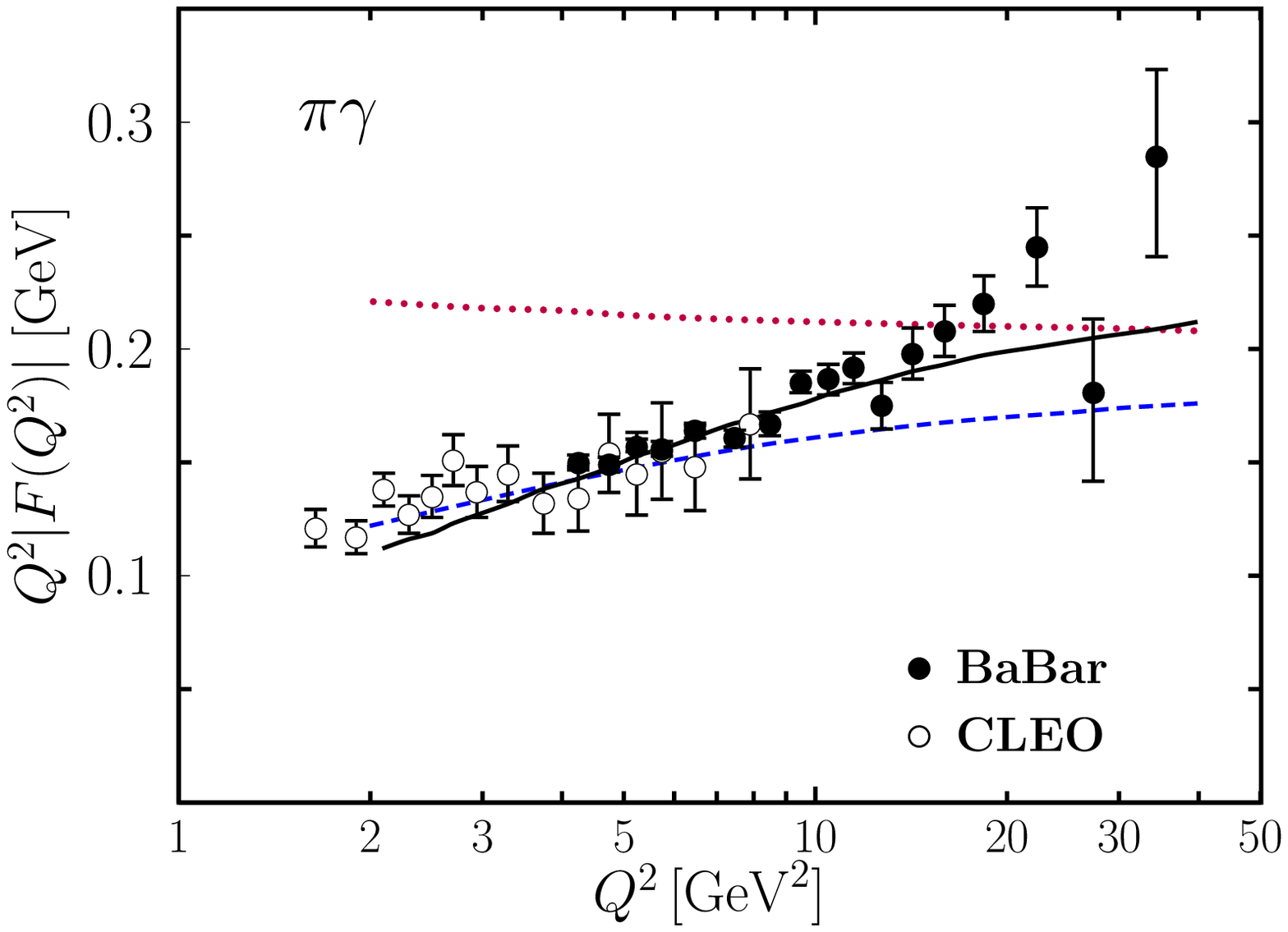} 
 \end{center}
\caption{The scaled $\pi\gamma$ transition form factor versus $Q^2$ evaluated
  from fit \req{eq:a4-fit} (solid line). The dashed line represents 
  the result given in \ci{raulfs95} which is obtained from the asymptotic 
  \da{} and $\sigma_\pi=0.861\,\gev^{-1}$ given. The dotted line is obtained 
  from collinear factorization to NLO accuracy. Data taken from 
  \ci{babar09,CLEO}. (Colors online)}
\label{fig:fit-a4}
\end{figure}
Two fits are ultimately performed: for the first one the Gegenbauer
coefficient $a_2$ is fitted; for the second one $a_4$ is freed but $a_2$ is
fixed at the face result of a recent lattice QCD calculation \ci{lattice}.
Evolved to the scale $\mu_0=2\,\gev$ with $\LQCD=181\,\mev$, the lattice 
result \ci{lattice} is $a_2(\mu_0)=0.201\pm 0.114$. All other Gegenbauer 
coefficients are assumed to be negligible in the fits. The results
of the two fits are~\footnote{
The Gegenbauer coefficient $a_n$ at a scale $\mu_1$ is obtained from the
value quoted in \req{eq:a2-fit} or \req{eq:a4-fit} by multiplying it with the
factor $[\ln{(\muO^2/\LQCD^2)}/\ln{(\mu_1^2/\LQCD^2)}]^{\gamma_n}$.}
\be
\sigma_\pi\=0.40\pm 0.06\,\gev^{-1}\,, \qquad a_2(\muO)\=0.22\pm 0.06\,, \quad
\chi^2=34.1\,,
\label{eq:a2-fit}
\ee
and with $a_2(\muO)=0.20$ \ci{lattice},
\be
\sigma_\pi\=0.40\pm 0.06\,\gev^{-1}\,, \qquad a_4(\muO)\=0.01\pm 0.06\,, \quad
\chi^2=34.2\,,
\label{eq:a4-fit}
\ee
The fits are stable and have a sharp  $\chi^2$ minimum. The minimal values of 
$\chi^2$ for the two fits are reasonable given that 28 data points for 
$Q^2>2.3\,\gev^2$ are included in the fits. The fit \req{eq:a2-fit} yields a 
value for $a_2$ that is in good agreement with the lattice result \ci{lattice}
within errors. The fit \req{eq:a4-fit} is shown in Fig.\ \ref{fig:fit-a4}; 
the fit \req{eq:a2-fit} is practically indistinguishable from \req{eq:a4-fit}. 
At the largest values of $Q^2$ the resultant form factor seems to be a bit 
small as compared to the BaBar data. Partially responsible for this fact are 
the large fluctuations the BaBar data exhibit; the fit compromises between all 
the data. At very large values of $Q^2$, not shown in Fig.\ \ref{fig:fit-a4}, 
the theoretical result for the form factor, scaled by $Q^2$, flattens off 
with a broad maximum at about $200\,\gev^2$ and decreases subsequently towards 
its asymptotic value of $\sqrt{2} f_\pi$. The reason for the small value of 
$\sigma_\pi$ in the present fits can easily be understood. The ${\cal C}_2$
and ${\cal C}_4$ terms demanded by the BaBar data \ci{babar09} at large $Q^2$, 
also contribute at low $Q^2$, although to a lesser extent. Therefore, the 
${\cal C}_0$-term with $\sigma_\pi=0.861\,\gev^{-1}$ which fits nicely the low 
$Q^2$ data, see Fig.\ \ref{fig:fit-a4}, is to be reduced. The only parameter 
available for this is however $\sigma_\pi$.

In Ref.\ \ci{BHL} it is argued that the decay $\pi^0\to\gamma\gamma$ fixes 
the \wf{} at $\vk=0$ integrated over $x$. For the \wf{} \req{eq:Gaussian} 
this constraint can be turned into a result for the transverse size 
parameter \ci{BKS}
\be
\sigma_\pi \= \Big[8 \pi^2 f_\pi^2 \sum_{n=0,2,4\ldots}\, a_n\Big]^{-1/2}\,.
\label{eq:BHL}
\ee
For the asymptotic \da{} this constraint leads to the value 
$\sigma_\pi=0.861\,\gev^{-1}$ used in \ci{raulfs95}. The parameters given in 
\req{eq:a2-fit}, \req{eq:a4-fit} violate \req{eq:BHL} plainly. Since it 
is however unclear at which scale the constraint \req{eq:BHL} holds the 
significance of this violation cannot be judged.

Also shown in Fig.\ \ref{fig:fit-a4} is a typical result of the collinear 
factorization approach to NLO accuracy (taking $\mu_F=\mu_R=Q$ and, as an
example, the Gegenbauer coefficients of the fit \req{eq:a4-fit}). Apparently 
the shape of that result is in conflict with experiment. A change of the
values of the Gegenbauer coefficients or the addition of further coefficients 
with the proviso that large negative coefficients are excluded, does not alter 
the shape but only the absolute value of the form factor. Thus, the 
$\pi\gamma$ transition form factor sets an example of a simple exclusive 
observable for which collinear factorization is insufficient for $Q^2$ as 
large as $40\,\gev^2$.    

It is perhaps of interest to examine the role of the soft intrinsic transverse
momentum or $b$-dependence of the \wf{} further. Thus, one may wonder whether 
or not a possible scale dependence of the transverse-size parameter improves
the fit. This scale dependence has been investigated in \ci{vogt} and found 
that $\sigma_\pi$ slowly decreases with increasing $Q^2$. Employing the
numerical results given in \ci{vogt} in the fit, one obtains results for the 
form factor that are very similar to the ones presented in Fig.\ \ref{fig:fit-a4}. 
As in most applications of light-cone \wf s an approximately scale independent 
transverse-size parameter is therefore assumed in this work for convenience.
If the transverse-size dependence in \req{eq:Gaussian} is neglected which 
leads to the original version of the MPA \ci{li92}, a food fit to the form 
factor data cannot be achieved, the results are too flat as compared to the
data and the minimal $\chi^2$ is 155. Hence, the $\vbs$ dependence of the \wf{}
is an important ingredient of the MPA as has been suggested in \ci{jakob93}.
A $b$-dependence of the \wf{} like
\be
\hat{\Psi}_P \propto \exp{[-\vbs^2/4\sigma_\pi^2]}
\label{eq:simple}
\ee
leads to fits of similar quality as with \req{eq:Gaussian}. The obtained
values of the Gegenbauer coefficients are slightly larger and have larger
errors. Within these errors they however agree with those quoted in 
\req{eq:a2-fit}, \req{eq:a4-fit} within one standard deviation. 

One may evaluate the probability \req{eq:probability} of the pion's valence
Fock state and the r.m.s.\ value of $\vk$ \req{eq:rms-k}. From the parameters 
quoted in \req{eq:a2-fit} and \req{eq:a4-fit} one finds $P_{q\bar{q}}\simeq 0.06$
and $\langle \vk^2\rangle^{1/2}\simeq 710\,\mev$. These values
appear to be plausible at a low scale although, in view of the assumption of a
scale-independent $\sigma_\pi$, they are to be taken with a grain of salt. 

\begin{figure}[th]
\begin{center}
 \includegraphics[width=.45\tw, bb=110 369 585 717,clip=true]
{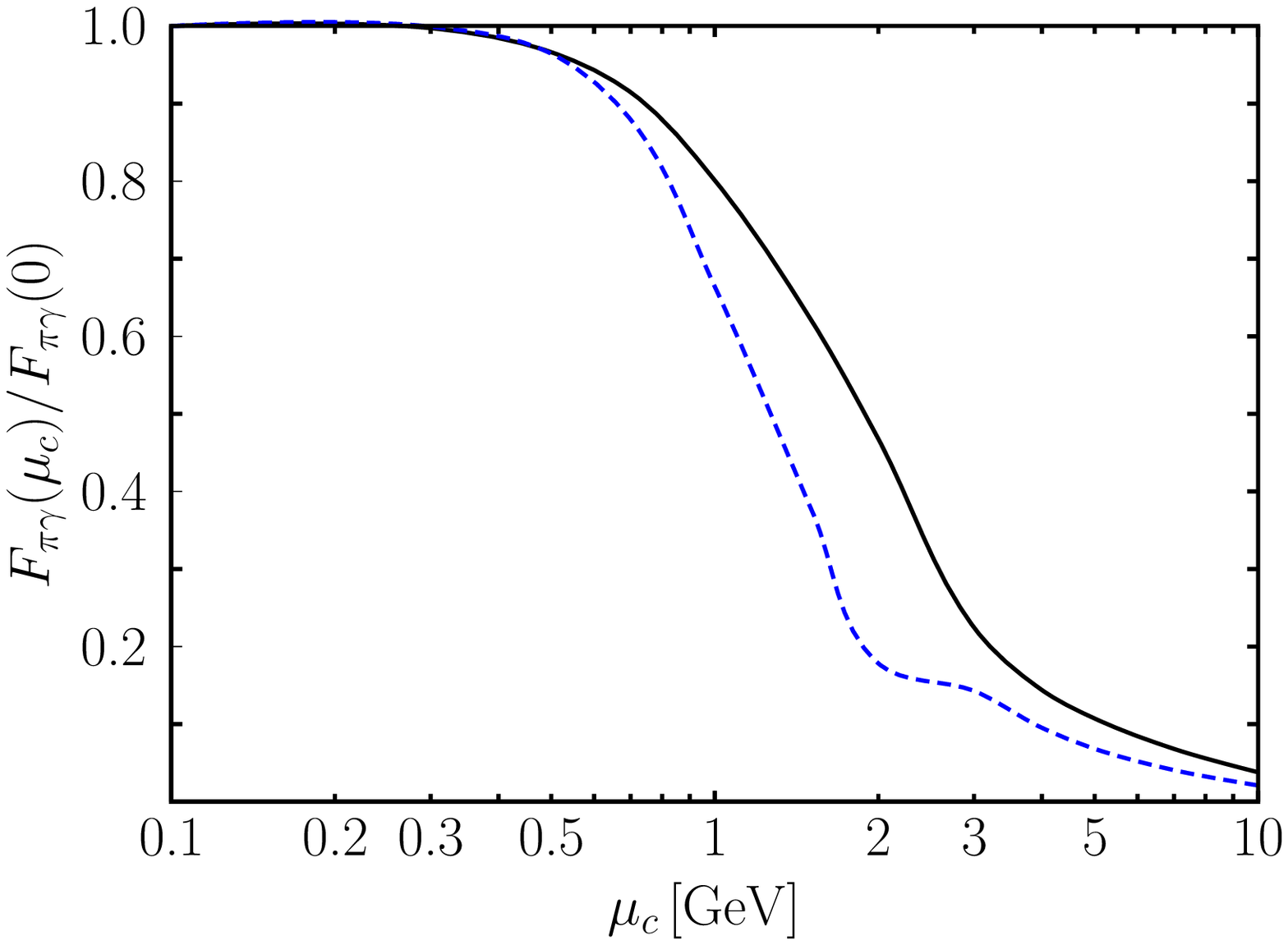} 
 \includegraphics[width=.45\tw, bb=110 380 585 730,clip=true]{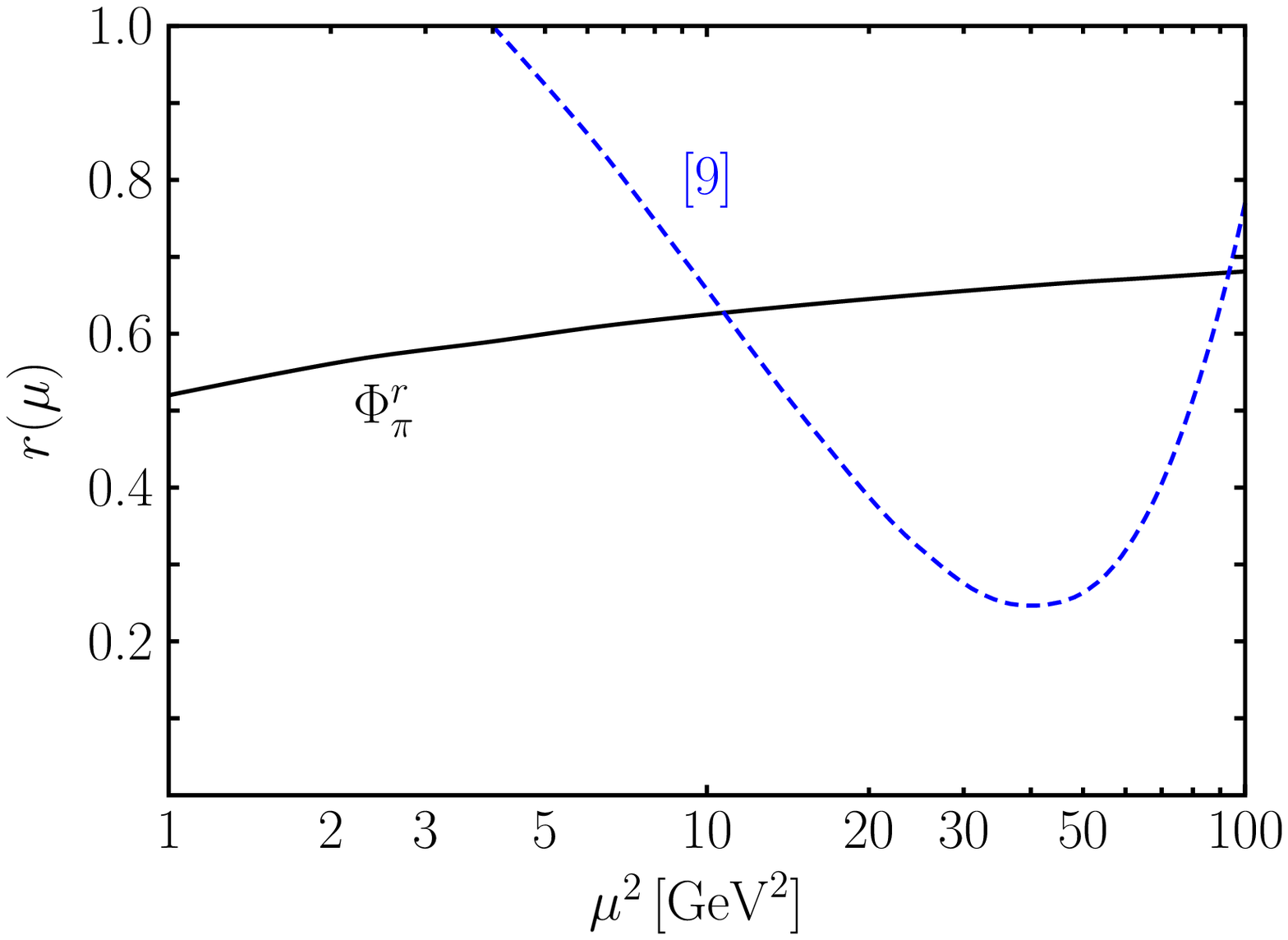} 
 \end{center}
\caption{Left: The accumulation profile of the $\pi\gamma$ transition form factor 
  in the space (solid line) and time-like (dashed line) regions versus a cut-off 
  of the renormalization scale at $Q^2=10\,\gev^2$ evaluated for fit
  \req{eq:a4-fit}. The absolute value of the time-like form factor is displayed. 
  (Colors online). Right: The effective power of the \da{} \req{eq:powerDA}
  compared to the power of the threshold factor in \ci{li09} at the scale $\mu$. }
\label{fig:accumulation}
\end{figure}

Since the integrations in \req{eq:FF-mpa} extend from $x=0$ to 1 and from 
$b=0$ to $1/\LQCD$  the soft regions contribute to the form factor too. 
The decisive question is how much. In order to investigate this issue a 
cut-off parameter $\mu_c$ is introduced and any contribution to the form factor
is set to zero if the renormalization scale \req{eq:renormalization} is less
than $\mu_c$. The accumulation profile is then defined by the ratio 
$F_{\pi\gamma}(\mu_c)/F_{\pi\gamma}(0)$. It is shown in Fig.\
\ref{fig:accumulation} and reveals that in the soft regions defined by small 
values of the renormalization scale \req{eq:renormalization}, say, less than 
$1\,\gev$, only small contributions are accumulated. The bulk of the
contributions to the transition form factor is generated in regions where the
renormalization scale is sufficiently large. Hence, within the MPA, the 
$\pi\gamma$ transition form factor can be considered as being calculated
self-consistently. This property holds down to values of the photon virtuality
of about $2\,\gev^2$.

Li and Mishima \ci{li09} also applied the MPA to the $\pi\gamma$ transition
form factor and achieved a reasonable fit to the CLEO \ci{CLEO}
and BaBar data \ci{babar09}. In contrast to the present work the flat \da{} \ci{rad09}
\be
\Phi_\pi\equiv 1
\label{eq:flat}
\ee
is used. It is combined with a Gaussian $b$-dependence as in \req{eq:Gaussian}
in a kind of wave function. However, this product cannot be considered as a
proper \wf{} in so far as it is not normalizable, see \req{eq:probability}~\footnote{
   The Gegenbauer series of the flat \da{} truncated at $n_0$ is however 
   normalizable provided $n_0$ is not too large. Alternatively one may use 
   \req{eq:simple}  which leads to a normalizable \wf{} and provides
   similar results for the transition form factor if the transverse size
   parameter is appropriately chosen.}. 
It is furthermore argued in \ci{li09} that the \da{} \req{eq:flat} is
accompanied by a threshold factor that represents resummed double logs 
$\als \ln^2{x}$ and $\als \ln^2{(1-x)}$ arising from the end-point
singularities which occur for the flat \da{} in collinear factorization. 
The threshold factor combined with the flat \da{} can be viewed as an 
effective \da{} of the type
\be
 \Phi_\pi^r\= \frac{\Gamma(2+2r)}{\Gamma^2(1+r)}\,\big[x\xb\big]^{r}\,.
\label{eq:powerDA} 
\ee 
According to \ci{li09}, $r$ is about 1 for low $Q^2$ and small for   
$Q^2\simeq 40\,\gev^2$. This particular $Q^2$-dependence of the power $r$
generates the increase of the form factor required by the BaBar data 
\ci{babar09}: At low $Q^2$ the effective \da{} is the asymptotic one 
implying small values of the transition form factor while, at large $Q^2$, 
the effective \da{} is close to the flat one and hence leads to much larger
values of the form factor. It is to be stressed that QCD evolution of the 
proper \da{}, $\Phi_\pi=1$, is omitted in \ci{li09}. Since the gliding 
factorization scale $\muF$ is an essential part of the MPA this seems to be 
a problematic assumption.

Eq.\ \req{eq:powerDA} defines a family of power-like \da s. It includes the
limiting cases of the asymptotic \da{} for $r=1$ as well as the flat \da{}
\req{eq:flat} for $r=0$. Also the square root \da{} proposed in \ci{bro07}
belongs to this family. By comparison with the Gegenbauer series 
\req{eq:distribution-amplitude} of \req{eq:powerDA} and its evolution one can 
show \ci{BDK} that the \da{} \req{eq:powerDA} approximately remains 
power-like under evolution, $r \to r(\mu)$ over a large range of the scale. 

One may examine the power-like \da{} \req{eq:powerDA} by fitting the
transverse size parameter as well as the power $r(\mu_0)$ to the data on the
$\pi\gamma$ form factor. One finds ($\muO=2\,\gev$) 
\be
\sigma_\pi= 0.40\pm 0.05\,\gev^{-1}\, \quad r(\muO)\=0.59\pm 0.06\,, 
                        \qquad \chi^2\=34.4\,. 
\label{eq:powers}
\ee
The quality of this fit to the data on the $\pi\gamma$ form factor is 
similar to that presented in \ci{li09}. In Fig.\ \ref{fig:accumulation} the
power $r(\mu)$ for the fit \req{eq:powers} is compared to the scale dependence
of the threshold factor used in \ci{li09} (in this work the power is set to
unity for $\mu^2\lsim 4\,\gev^2$). As can be seen from Fig.\
\ref{fig:accumulation} the \da{} \req{eq:powerDA} exhibits the usual evolution
behavior, it monotonically evolves into the asymptotic one, $\Phi_{\rm AS}$,
for $\mu\to\infty$. On the other hand, the scale dependence of the threshold
factor or the effective \da{} advocated for in \ci{li09}, is drastically
different. It is also to be stressed that in \ci{li09} the threshold factor is 
evaluated at the scale $Q$ and not at the factorization scale $\mu_F$ although
it is to be understood as part of the \wf{}.

The first few Gegenbauer coefficients of the power-like \da{} \req{eq:powers}
are compared to those from the fits \req{eq:a2-fit} and \req{eq:a4-fit} in
Tab.\ \ref{tab:1}. The first Gegenbauer coefficients of the flat and the 
square root \da{} advocated for in \ci{bro07}, are also displayed in Tab.\ 
\ref{tab:1} for comparison. The comparison is made at a scale of $1\,\gev$ 
which seems to be a plausible value for soft \wf s like \req{eq:flat}.
While the Gegenbauer coefficients of the flat \da{} are generally larger 
than those quoted in \req{eq:a2-fit} or \req{eq:a4-fit}, are the coefficients 
of \req{eq:powers} and the square root \da{} smaller. A \wf{} like 
$\Psi=c\,\exp{[-\vbs^2/4\sigma_\pi^2]}$ to which the flat \da{} is associated, 
can be used in a MPA calculation without entailing infrared singularities 
even if evolution is ignored. On the other hand, the use of \req{eq:flat} 
within the collinear factorization approach leads to an infrared-singular 
result which necessitates a regularization prescription as for instance 
the insertion of a mass parameter into the denominator of the quark 
propagator \ci{pol09}. Such a regularization prescription typically alters 
the asymptotic behavior of the scaled form factor drastically: instead of 
$Q^2\,F_{\pi\gamma}\to$ const. the scaled form factor increases 
logarithmically \ci{rad09,pol09,dorokhov10}.
\begin{table*}[t]
\renewcommand{\arraystretch}{1.4} 
\begin{center}
\begin{tabular}{| c || c |c | c |}
\hline     
      & $a_2$  & $a_4$ & $a_6$ \\[0.2em]
\hline
\req{eq:a2-fit}  & $0.27\pm 0.07$ & 0 & 0 \\[0.2em]
\req{eq:a4-fit}  & 0.25 & $0.01\pm 0.06$ & 0 \\[0.2em]
\req{eq:powers}  & 0.14 & 0.05 & 0.03 \\[0.2em]
\req{eq:flat} & 0.39 & 0.24 & 0.18 \\[0.2em]
$\sqrt{x\xb}$  & 0.15 & 0.06 & 0.03 \\[0.2em]
\hline  
\end{tabular}
\end{center}
\caption{Lowest Gegenbauer coefficients for various \da{} at the scale $1\,\gev$.}
\label{tab:1}
\renewcommand{\arraystretch}{1.0}   
\end{table*}  

\section{Generalization to the $\eta$ and $\eta^\prime\gamma$ 
           form factors}
\label{sec:eta}
The analysis of the $\pi\gamma$ transition form factor within the MPA can
straightforwardedly be extended to the cases of the $\eta\gamma$ and
$\eta^\prime\gamma$ ones \ci{feldmann97,uppsala}. 
These transition form factors may be expressed as a
sum of the flavor-octet and flavor-singlet contributions ($P=\eta, \eta^\prime$)
\be
F_{P\gamma} \= F^8_{P\gamma} +  F^1_{P\gamma}\,.
\ee
As is the case for the $\pi\gamma$ form factor the functions $F^i_{P\gamma}$
($i=1, 8$) are proportional to the constants $f_P^i$ assigned to the decays 
of meson $P$ through the SU(3)$_F$ octet or singlet axial-vector weak currents 
which are defined by the matrix elements
\be
\langle 0\mid J_{\mu 5}^i\mid P(p)\rangle \= i f_P^i\, p_\mu  
\ee
Adopting the general parameterization \ci{kaiser}
\ba
f_\eta^8 &=& f_8 \cos{\theta_8}\,, \qquad f_\eta^1 \=- f_1 \sin{\theta_1}\,,\nn\\
f_{\eta^\prime}^8 &=& f_8 \sin{\theta_8}\,, \qquad 
f_{\eta^\prime}^1 \=\phantom{-} f_1 \cos{\theta_1}\,,
\ea
one can show \ci{FKS1} that on exploiting the divergences of the axial-vector
currents - which embody the axial-vector anomaly - the mixing angles,
$\theta_8$ and $\theta_1$, differ considerably from each other and from the
state mixing angle, $\theta$. In \ci{FKS1} the mixing parameters have
been determined:
\ba
f_8&=&1.26\, f_\pi\,, \qquad  f_1\=1.17\, f_\pi\,, \nn\\
\theta_8&=&-21.2^\circ\,, \qquad \theta_1\=-9.2^\circ\,.
\label{eq:mixing-par}
\ea 

Assuming particle-independent \ci{feldmann97,FKS1} octet and singlet \wf s, 
$\Psi^{8}$ and $\Psi^{1}$ , which are parameterized as in \req{eq:Gaussian}
with the respective decay constants $f_8$ and $f_1$ instead of $f_\pi$,  one
can cast the transition form factors into the form
\ba
F_{\eta\gamma} &=& \cos{\theta_8}\,F^8 - \sin{\theta_1}\,F^1\nn\\
F_{\eta^\prime\gamma} &=& \sin{\theta_8}\,F^8 + \cos{\theta_1}\,F^1\,.
\label{FF-mixing}
\ea 
The charge factors in \req{eq:th} read (with $P=1,8$)
\be
C_8\=(e_u^2 + e_d^2 -2e_s^2)/\sqrt{6}\,, \qquad C_1\=(e_u^2 + e_d^2 + e_s^2)/\sqrt{3}\,.
\ee 
The asymptotic behavior of the form factors is
\be
Q^2 F^8 \to \sqrt{\frac23} f_8\,, \qquad Q^2 F^1 \to \frac{4}{\sqrt{3}} f_1\,.
\label{octet-singlet-asymptotics}
\ee
It is to be noted that the singlet-decay constant is renormalization-scale
dependent \ci{kaiser}:
\be
\mu \frac{df_1}{d\mu} \= \gamma_A(\mu)\,f_1
\ee
where the anomalous dimension is
\be
\gamma_A\= -n_f\left(\als(\mu)/\pi\right)^2\,.
\ee
Since the anomalous dimension controlling this scale dependence is of order 
$\als^2$, it leads to tiny effects. In fact, if the value of $f_1$ quoted 
in \req{eq:mixing-par} is to be understood as being valid at, say, the  
scale $\simeq 1\,\gev$, its asymptotic value is $13\%$ smaller; over the 
range of available data it only decreases by less than $2\%$. The scale 
dependence of $f_1$ is therefore discarded for convenience. 

It is to be stressed that to NLO of the hard scattering there is also a
contribution from the glue-glue Fock component of the mesons  to the singlet
form factor \ci{grozin,passek}. In the MPA analysis performed in this work 
the glue-glue Fock component does not contribute directly but only through 
the matrix of the anomalous dimensions and the  mixing of the singlet 
quark-antiquark with the glue-glue \da{}. It is assumed here that the
Gegenbauer coefficients of the glue-glue distribution amplitude are zero at
a low scale of order $1\,\gev$. Hence, the quark-antiquark distribution 
amplitude evolves with the eigenvalues $\gamma_n^{(+)}$ of the anomalous 
dimension matrix \ci{passek}. The values of $\gamma_n^{(+)}$ practically fall 
together with those of $\gamma_n$, the anomalous dimension of the octet 
distribution amplitude, see \req{eq:anomalous-dimensions}. For instance, 
$\gamma_2^{(+)}=0.639$ while $\gamma_2=2/3$ for four flavors.

The two form factors $F^8$ and $F^1$ can now be evaluated from \req{eq:FF-mpa} 
and \req{eq:th} in full analogy to the $\pi\gamma$ transition form factor. 
The data on $F^8$ and $F^1$ are extracted from the CLEO \ci{CLEO} and BaBar 
\ci{BaBar-prel} data using \req{FF-mixing}. As for the $\pi\gamma$ form 
factor only the transverse size parameter and one Gegenbauer coefficient for 
each \wf{} can be determined. The best fit is obtained with the parameters 
($\muO=2\,\gev$): 
\ba
\sigma_8&=& 0.84\pm 0.14\,\gev^{-1}\,, \qquad  a_{2}^8(\muO)\= -0.06\pm 0.06\,, \nn\\
\sigma_1&=& 0.74\pm 0.05 \,\gev^{-1}\,, 
                     \qquad a_{2}^1(\muO)\= -0.07\pm 0.04\,.
\label{eta-fit}
\ea
The values of $\chi^2$ are 15.0 and 14.1 for the octet and singlet
cases, respectively (for 16 data points in each case). The probabilities of
the singlet and octet \wf s are 0.24 and 0.19, respectivly. The corresponding
r.m.s. $\vk$-values are 390 and $440\,\mev$. 

\begin{figure}[t]
\begin{center}
  \includegraphics[width=.50\tw, bb=105 370 571 718,clip=true]
{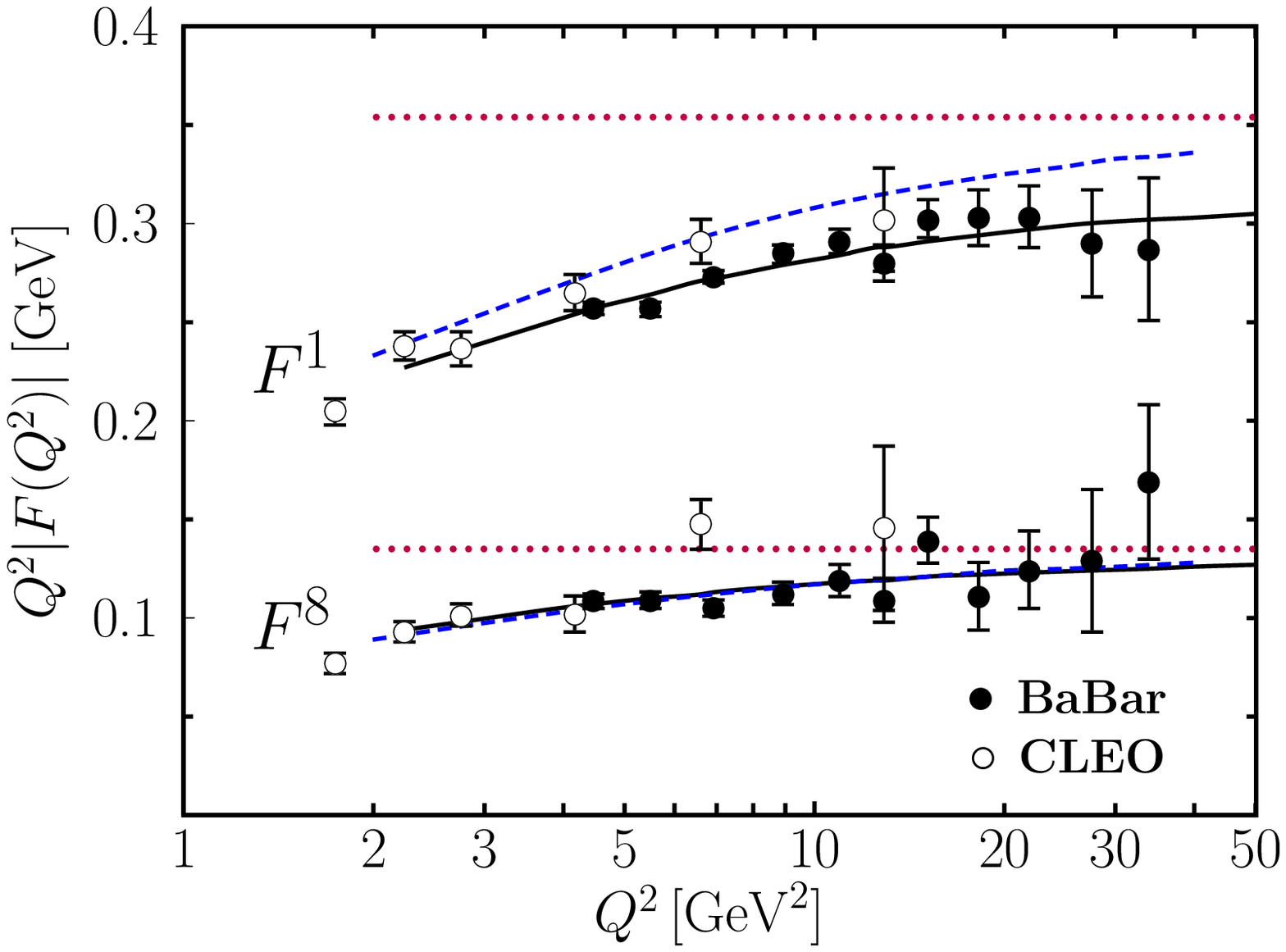}
  \end{center}
\caption{The octet and singlet form factors. Dotted lines represent the
  asymptotic behavior \req{octet-singlet-asymptotics}, the dashed lines 
  the results obtained in \ci{feldmann97}. The solid lines represent the 
  new fit \req{eta-fit}. Data taken from \ci{CLEO,BaBar-prel}. (Colors online)}
\label{fig:singlet-octet}
\end{figure}

\begin{figure}[t]
\begin{center}
  \includegraphics[width=.47\tw, bb=106 366 570 709,clip=true]
                                       {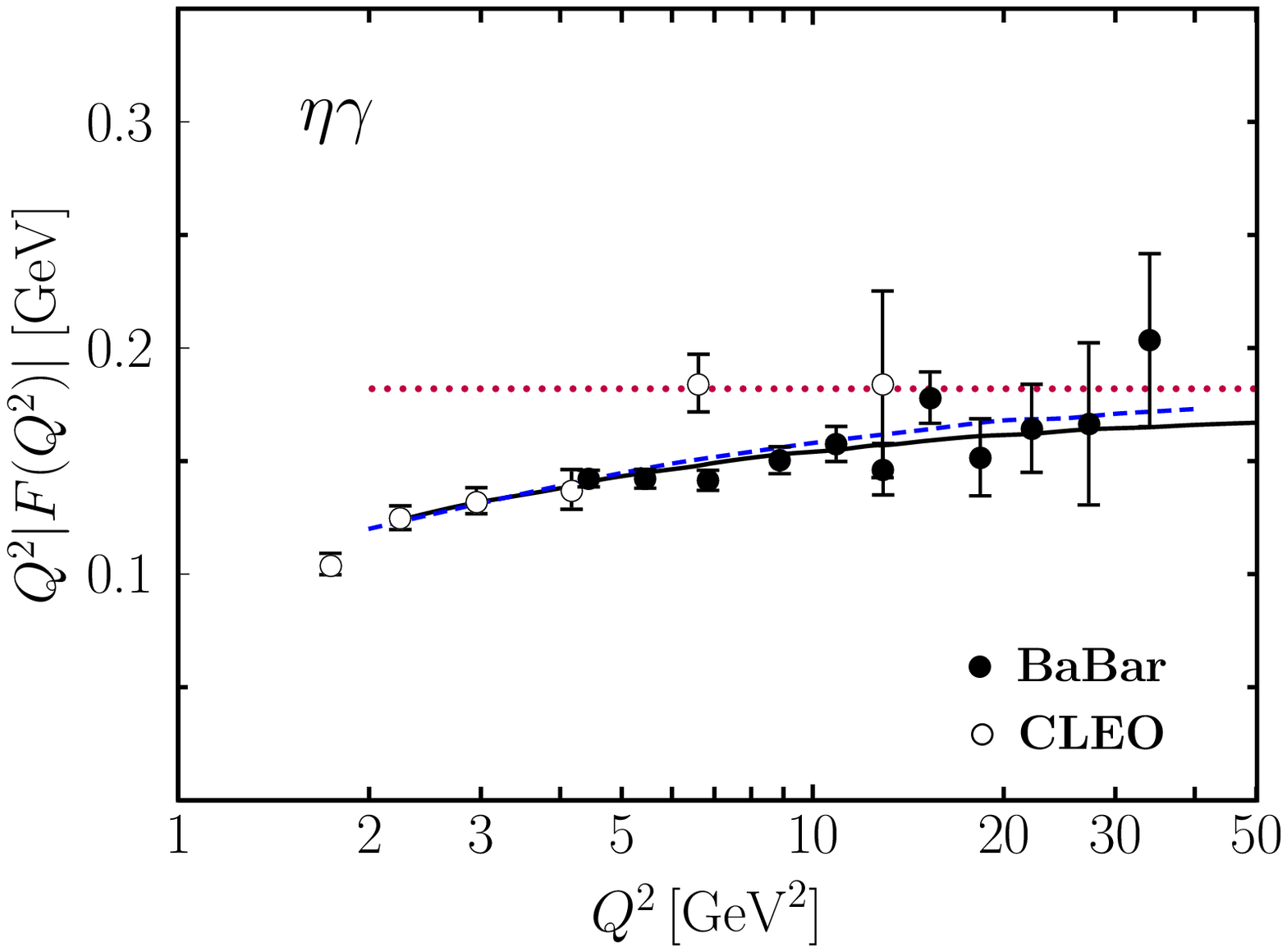}
 \includegraphics[width=.47\tw, bb=127 373 592 718,clip=true]
                                      {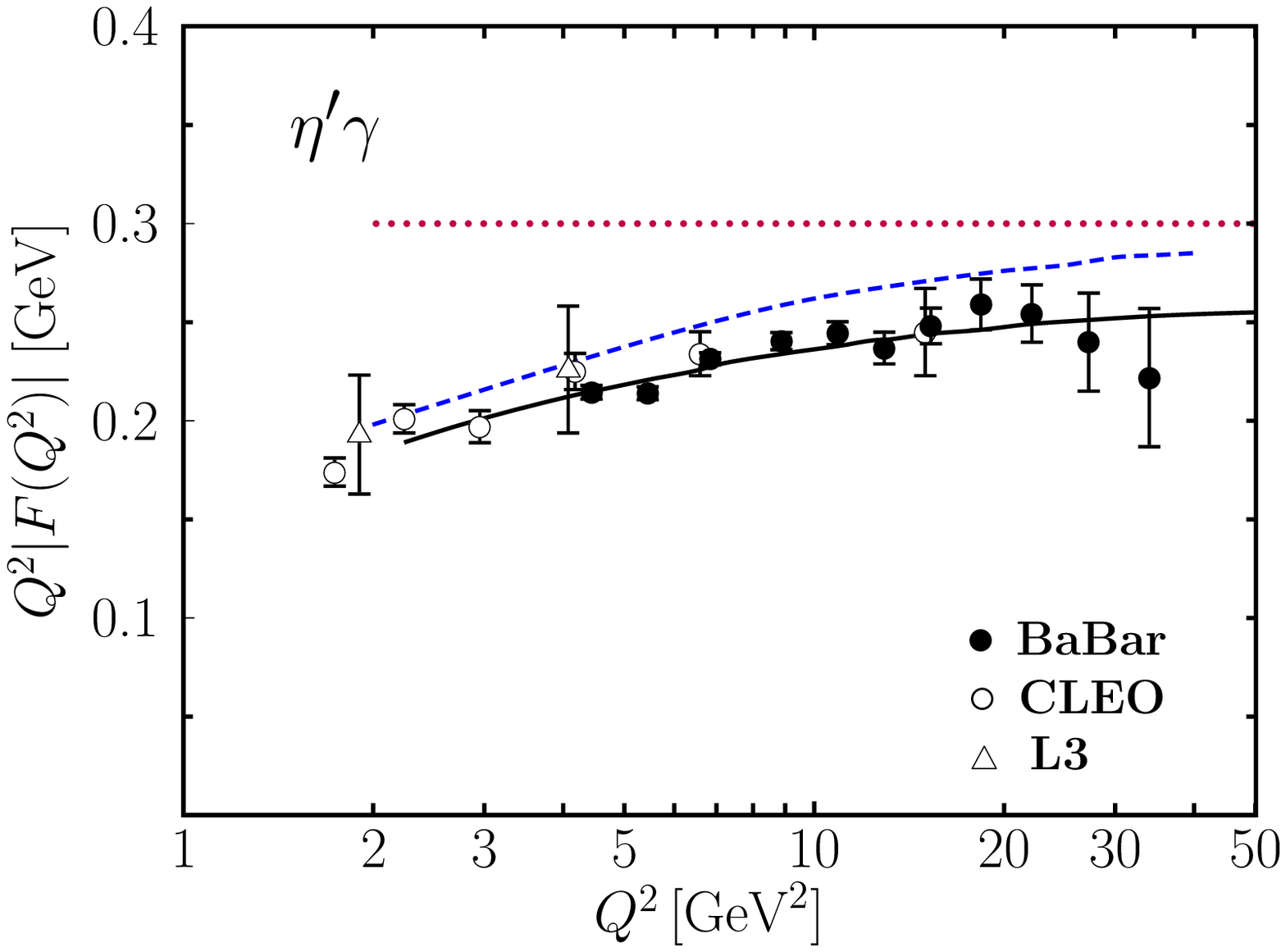}
 \end{center}
\caption{The scaled $\eta\gamma$ and $\eta^\prime\gamma$ transition form
  factor versus $Q^2$. Data taken from \ci{CLEO,L3,BaBar-prel}.
  For notations refer to Fig.\ 
  \ref{fig:singlet-octet}. (Colors online)}
\label{fig:eta-etap}
\end{figure}

In Fig.\ \ref{fig:singlet-octet} the results of this fit are compared to the
data on $F^8$ and $F^1$. The quality of this fit is very good. In contrast to 
the $\pi\gamma$ case the data on both $F^8$ and $F^1$ lie below the asymptotic 
results \req{octet-singlet-asymptotics}.
The combination of these two form factors into the physical ones according to 
\req{FF-mixing} leads to the results which are shown in Fig.\
\ref{fig:eta-etap}. Again very good agreement with the data is to be observed. 
Also shown in Figs.\
\ref{fig:singlet-octet} and \ref{fig:eta-etap} are the results obtained in 
\ci{feldmann97} which have been evaluated from the asymptotic distribution
amplitudes (with $\sigma_1=\sigma_{8}=0.861\,\gev^{-1}$). The octet as well as
the $\eta\gamma$ form factors of \ci{feldmann97} are in very good agreement
with experiment while the results for $F^1$ and the $\eta^\prime\gamma$ form
factor are somewhat too large. 

\begin{figure}[ht]
\begin{center}
\includegraphics[width=.45\tw, bb=102 380 574 727,clip=true]
                                   {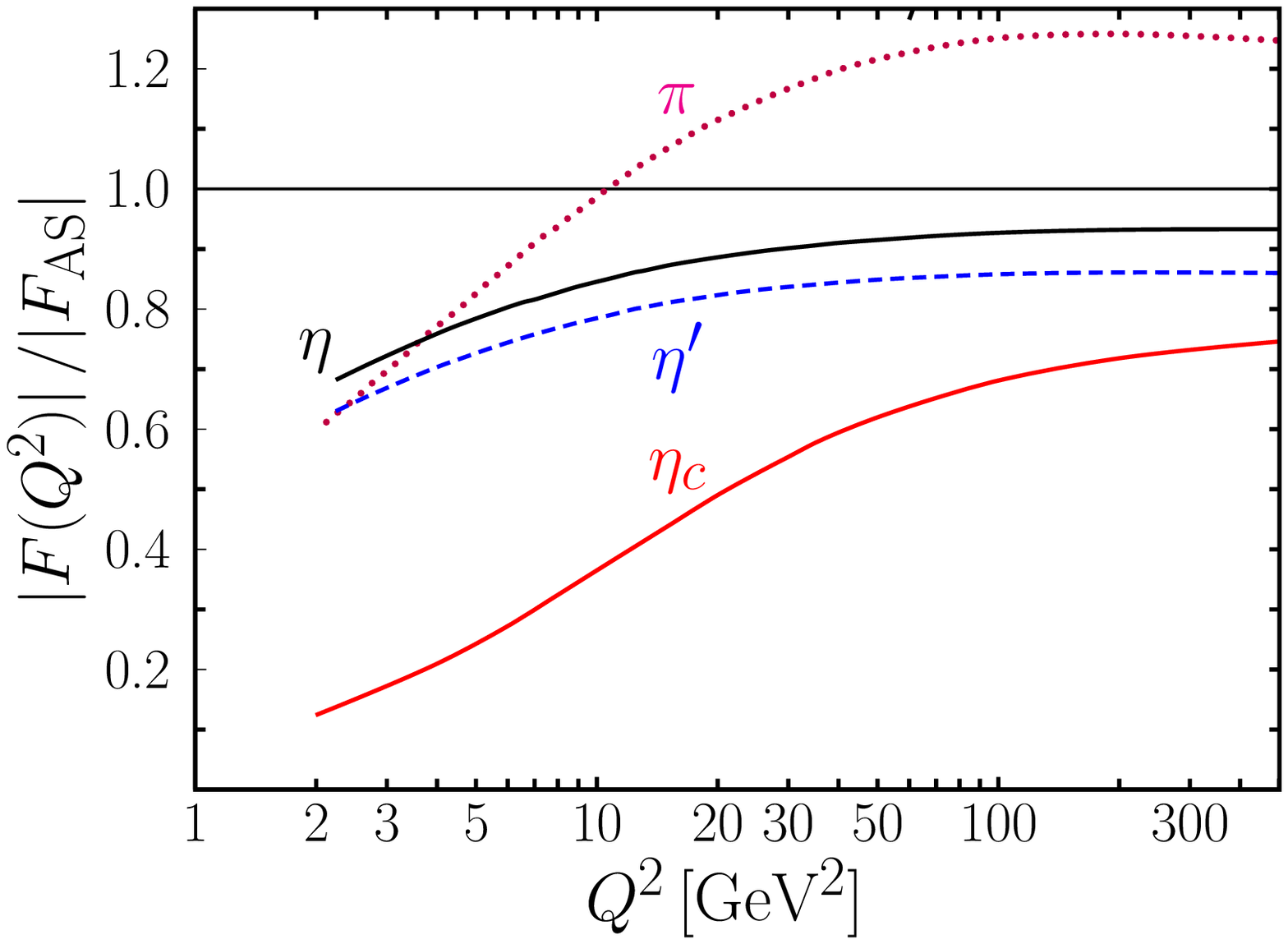}
\includegraphics[width=.45\tw, bb=100 351 569 701,clip=true]{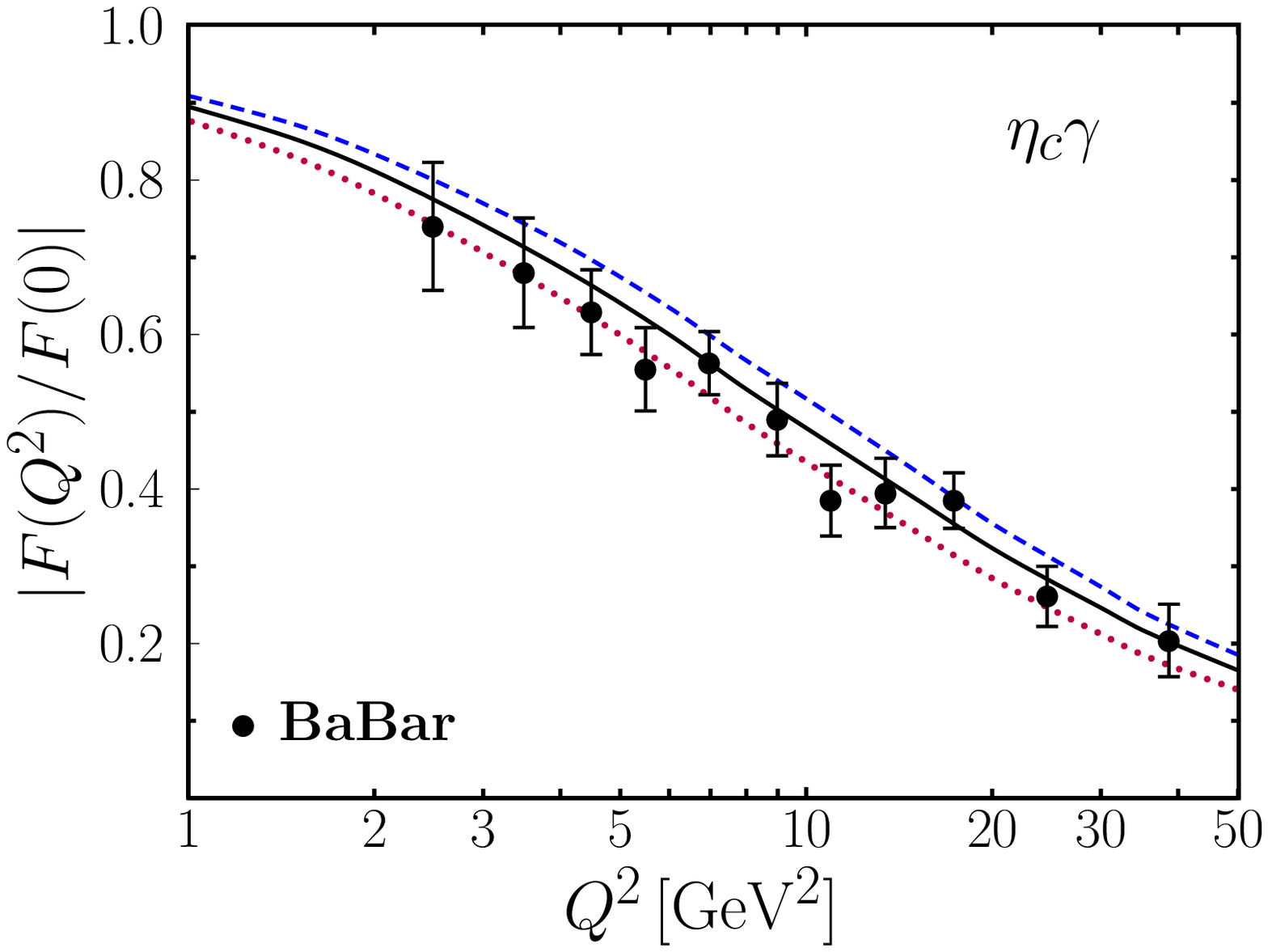}
 \end{center}
\caption{Left: The $P-\gamma$ transition form factors scaled by the
  corresponding asymptotic behavior, versus $Q^2$. The thick solid (dashed,
  dotted, thin solid) line represents the case of the $\eta$ ($\eta^\prime$,
  $\pi$, $\eta_c$) meson. Parameters are taken from fit \req{eq:a4-fit}, 
  \req{eta-fit} and \req{eq:etac-wf}. Right: The $\eta_c\gamma$ form factor 
  scaled by its value at $Q^2=0$. Data taken from \ci{babar10}. The solid 
  (dotted) line represents the results of a calculation with the values of 
  the parameters quoted in \req{eq:etac-wf} (with $m_c=1.26\,\gev$). The 
  dashed line is the prediction given in \ci{feldmann97b}. (Colors online)}
\label{fig:comp}
\end{figure}

The form factors scaled by their respective asymptotic behaviors are displayed
in Fig.\ \ref{fig:comp}. All the form factor ratios for the light mesons  behave 
similar although not identical. Asymptotically they all tend to 1. The 
$\pi\gamma$ form factor approaches 1 from above, the other ones from
below. The approach to 1 is a very slow process; even at $500\,\gev^2$ the
limiting behavior has not yet been reached. It is also evident from Fig.\
\ref{fig:comp} that, forced by the BaBar data, there are strong violations of
SU(3)$_F$ flavor symmetry in the groundstate octet of the pseudoscalar mesons
at large $Q^2$. The difference between the $\pi\gamma$ and the $\eta\gamma$
or more precisely the $\eta_8\gamma$ form factors is larger than the
difference between their respective decay constants. In other processes
involving pseudoscalar mesons, e.g.\ two-photon annihilations into pairs 
of pseudoscalar mesons \ci{DK09,BELLE}, such large flavor symmetry violations 
have not been observed. Below $8\,\gev^2$, i.e.\ in the range of the CLEO
data, flavor symmetry breaking is much milder. The $\eta_c\gamma$ transition 
form factor which is also shown in Fig.\ \ref{fig:comp}, behaves very 
different. This form factor will be discussed in the next section.

\section{The case of the $\eta_c$}
\label{sec:etac}
The essential difference of the $\eta_c\gamma$ transition form factor to the 
other three form factors is the large mass of the $\eta_c$ ($M_{\eta_c}$) or 
that of the charm quark ($m_c$). They constitute a second large scale in
addition to the virtuality of one of the photons. The masses cannot be
neglected in a perturbative calculation in constrast to the case of the light 
mesons where quark and hadron masses do not play a role. Despite this the 
$\eta_c\gamma$ form factor is to be calculated from \req{eq:FF-mpa} but the 
hard scattering amplitude to lowest order perturbative QCD reads \ci{feldmann97b}
\be
T_H \= \frac{4\sqrt{6}\, e_c^2}{xQ^2 + (1+4 x\xb) m_c^2+\vk^2}\,.
\label{eq:etac-hsa}
\ee
The symmetry of the problem under the replacement of $x$ by $\xb$ is
already taken into account in \req{eq:etac-hsa}. Due to the involved second 
large scale in the problem the $\eta_c\gamma$ form factor can be calculated 
even at $Q^2=0$.\\
 
The light-cone \wf{} of the $\eta_c$ is parameterized as in \req{eq:Gaussian}~\footnote{
In \ci{feldmann97b} the Gaussian \req{eq:simple} is taken. The version 
\req{eq:Gaussian} is chosen here in order to be conform with the 
calculations of the other form factors. The differences between the two 
versions are marginal as has been mentioned previously.}. 
Following \ci{feldmann97b,bsw} the \da{} is chosen as  
\be
\Phi_{\eta_c} \= N(\sigma_{\eta_c}) x\xb 
   \exp{\Big[- \sigma^2_{\eta_c}M_{\eta_c}^2\frac{(x-1/2)^2}{x\xb}\Big]}
\ee
where $N(\sigma_{\eta_c})$ is determined from the usual requirement $\int_0^1
dx \Phi_{\eta_c}(x)=1$. The \da{} exhibits a pronounced maximum at $x=1/2$ and
is exponentially damped in the end-point regions. It describes an essentially 
non-relativistic $c\bar{c}$ bound state; quark and antiquark approximately 
share the meson's momentum equally. In the hard scattering
amplitude the charm quark mass occurs while in the \da{} the meson mass is
used. This property of the latter \da{} is a model assumption which
contributes to the theoretical uncertainty of the results. In the sense of 
the non-relativistic QCD \ci{HQET} $2m_c$ and $M_{\eta_c}$ are equivalent. 
In \req{eq:FF-mpa} the Sudakov factor $\exp{[-S]}$ can be set to 1 in the 
case at hand for two reasons: First, due to the large, non-negligible 
$c$-quark mass the radiative QCD corrections only produce soft divergencies 
but no collinear ones and, hence, the double logs do not appear. Second, 
the Sudakov suppressions is mainly active in the end-point regions (c.f.\ 
the discussion in Sect.\ \ref{sec:mpa}) which are already strongly damped 
by the $\eta_c$ \wf. The evolution behavior of the $\eta_c$ \da{} is unknown 
in the range where $Q^2$ is of order of $M_{\eta_c}^2$ and is therefore 
ignored here. Consequently, also the running of the charm quark mass is 
omitted. It has been checked that the effect of the this scale dependence 
is anyway only on the percent level.

The normalization of the $\eta_c\gamma$ transition form factor is fixed 
by its value at $Q^2=0$ which is related to two-photon decay width by
\be
\Gamma[\eta_c\to \gamma\gamma]\= \frac14\,\pi\ale^2\,M_{\eta_c}^3\,
                                    \mid F_{\eta_c\gamma}(0)\mid^2\,. 
\label{eq:etac-gamma-gamma}
\ee
However, this decay width is experimentally not well known \ci{PDG}. It is 
therefore advisable to normalize the form factor by its value at $Q^2=0$ all
the more so since the recent BaBar data \ci{babar10} are also presented this
way. Doing so the perturbative QCD corrections at $Q^2=0$ to the
$\eta_c\gamma$ transition form factor which are known to be large
\ci{barbieri}, are automatically included. Also the  $\als$ corrections for 
$Q^2\lsim M_{\eta_c}^2$ \ci{shifman} cancel to a high degree in the ratio 
$F_{\eta_c\gamma}(Q^2)/F_{\eta_c\gamma}(0)$. Even at $Q^2=10\,\gev^2$ their
effect is less than $5\%$, c.f.\ the discussion in \ci{feldmann97b}. 
The uncertainties in the present knowledge of the $\eta_c$ decay constant 
do also not enter the predictions for this ratio.

The recent Babar data on the $\eta_c\gamma$ form factor \ci{babar10} are shown
in the right hand panel of Fig.\ \ref{fig:comp}. The behavior of this data  
has indeed been predicted in \ci{feldmann97b}. The predictions which have been 
evaluated from $m_c=M_{\eta_c}/2$, are about one standard deviation too large 
but with regard to the uncertainties of the theoretical calculation, as for 
instance the exact value of the mass of the charm quark, one can claim 
reasonable agreement between theory and experiment. A little readjustment of 
the value of the charm quark mass improves the fit. Thus, with the parameters   
\be
m_c \=1.35\,\gev, \qquad \sigma_{\eta_c}\=0.44\,\gev^{-1}\,,
\label{eq:etac-wf}
\ee
a perfect agreement with experiment is achieved as is to be seen in Fig.\
\ref{fig:comp}. For comparison there are also shown results  evaluated from 
$m_c=1.21\,\gev$ in Fig.\ \ref{fig:comp}. As one may note from the left hand 
panel of Fig.\ \ref{fig:comp} the $\eta_c\gamma$ transition form factor 
behaves quite differently from the other three form factors. The large 
charm-quark mass slows down the approach to the asymptotic limit
\be
Q^2 F_{\eta_c\gamma} \to \frac{8f_{\eta_c}}{3}\,.
\ee  

Finally one may examine whether the parameters quoted in \req{eq:etac-wf} are
plausible. For this purpose the $\eta_c\gamma$ form factor at zero momentum 
transfer is evaluated. Ignoring $\als$-corrections as well as relativistic 
effects and using the value $420\,\mev$ for the $\eta_c$ decay constant 
$f_{\eta_c}$, one obtains
\be 
F_{\eta_c\gamma}(0)\= 0.085\,\gev^{-1}\,,
\ee
and a two-photon decay width \req{eq:etac-gamma-gamma} of
\be
\Gamma(\eta_c\to\gamma\gamma)\= 8.05\,\kev\,.
\label{eq:forward}
\ee
This result is in good agreement with the value of $(7.20\pm 2.11)\,\kev$
evaluated by the PDG \ci{PDG} and with recent theoretical estimates, see for
instance \ci{chao} and references therein. The parameters quoted in
\req{eq:etac-wf} together with $f_{\eta_c}\= 420\,\mev$ correspond to a 
normalization of the $\eta_c$ \da{} $N(\sigma_{\eta_c})\=8.849$, to a 
probability of the valence Fock state of 0.82 and to a r.m.s. $\vk$ of 
$773\,\mev$. The latter two values appear reasonable for a quarkonium state. 
 
\section{Two virtual photons}
\label{sec:two}
An extension of the MPA analysis of the $P\gamma$ transition form factors 
to the case of two virtual photons can be found in \ci{ong,DKV1,melic}. As 
an example the $\pi\gamma^*$ form factor will be discussed here. Denoting 
the photon virtualities by $Q^2$ and $Q^{\prime\, 2}$ and introducing the 
variables
\be
\ov{Q}^2\= \frac12\,(Q^2 + Q^{\prime\, 2})\,, \qquad 
\omega \= \frac{Q^2 - Q^{\prime\, 2}}{Q^2 + Q^{\prime\, 2}}\,,
\ee
one finds for the hard scattering amplitude the expression
\be
\hat{T}_H\= \frac{2}{\sqrt{3}\pi}\, K_0(\sqrt{1-\omega (1-2x)}\,\ov{Q}b)
\ee
in $b$-space. In generalization of \req{eq:FF-mpa} the $\pi\gamma^*$ form
factor now reads
\be
F_{\pi\gamma^*}(\ov{Q}^2,\omega) \= \int dx \frac{d^2\vbs}{4\pi}\,
\hat{\Psi}_\pi(x,-\vbs,\muF) \hat{T}_H(x,\vbs,\ov{Q},\omega,\muR) 
                                e^{-S(x,b,\ov{Q},\muF,\muR)}\,.
\label{eq:pi-gamma*}
\ee
The renormalization scale is taken as
\be
\muR\= \max{(\sqrt{1-\omega (1-2x)} \ov{Q},1/b)}
\ee
in generalization of \req{eq:renormalization}.

The form factor only falls off like $1/\ov{Q}^2$ at large $\ov{Q}^2$ in contrast
to the $Q^{-2} Q^{\prime\, -2}\propto \bar{Q}^{-4}$ behavior of the vector meson
dominance model \ci{kessler}. It is also interesting to note that like in
collinear factorization, the $\pi\gamma^*$ form factor is insensitive to the
higher Gegenbauer terms for $Q^2\simeq Q^{\prime 2}$; in the limit $\omega\to
0$ it only depends on the asymptotic \da{}. In fact it has been shown 
\ci{DKV1,melic} that the $n$-th order Gegenbauer term contributes to order 
$\omega^n$. The importance of effects from the intrinsic $\vk$ diminishes as 
both photons become virtual. This is evident from the quark propagator
in momentum space reading
\be
     T_H \propto [(1-\omega (1-2x)) \ov{Q}^2 + \vk^2]^{-1}\,. 
\label{eq:two-photon-propator}
\ee
As $\omega$ deviates from 1, the real photon limit, the form factor
$F_{\pi\gamma^*}$ becomes less sensitive to the end-point regions where either
the quark or the antiquark becomes soft. In the limit $\omega\to 0$ 
\req{eq:two-photon-propator} reduces to $[\ov{Q}^2+\vk^2]^{-1}$. This is to 
be contrasted with the $\omega\to 1$ limit in which 
$T_H\propto [xQ^2+\vk^2]^{-1}$, see \req{eq:hard-amplitude}. Thus, $\vk$ 
plays a minor role in the quark propagator provided $\omega$ is sufficiently 
small. In the limit $\omega\to 0$ and large $\ov{Q}^2$ the form factor becomes
\be
\ov{Q}^2 F_{\pi\gamma^*}(\ov{Q}^2,\omega) \= \frac{\sqrt{2}}3 f_\pi\,,
\ee
a result that has been derived long ago in \ci{chase}. 

\begin{figure}[ht]
\begin{center}
\includegraphics[width=.475\tw, bb=105 370 582 708,clip=true]
                                  {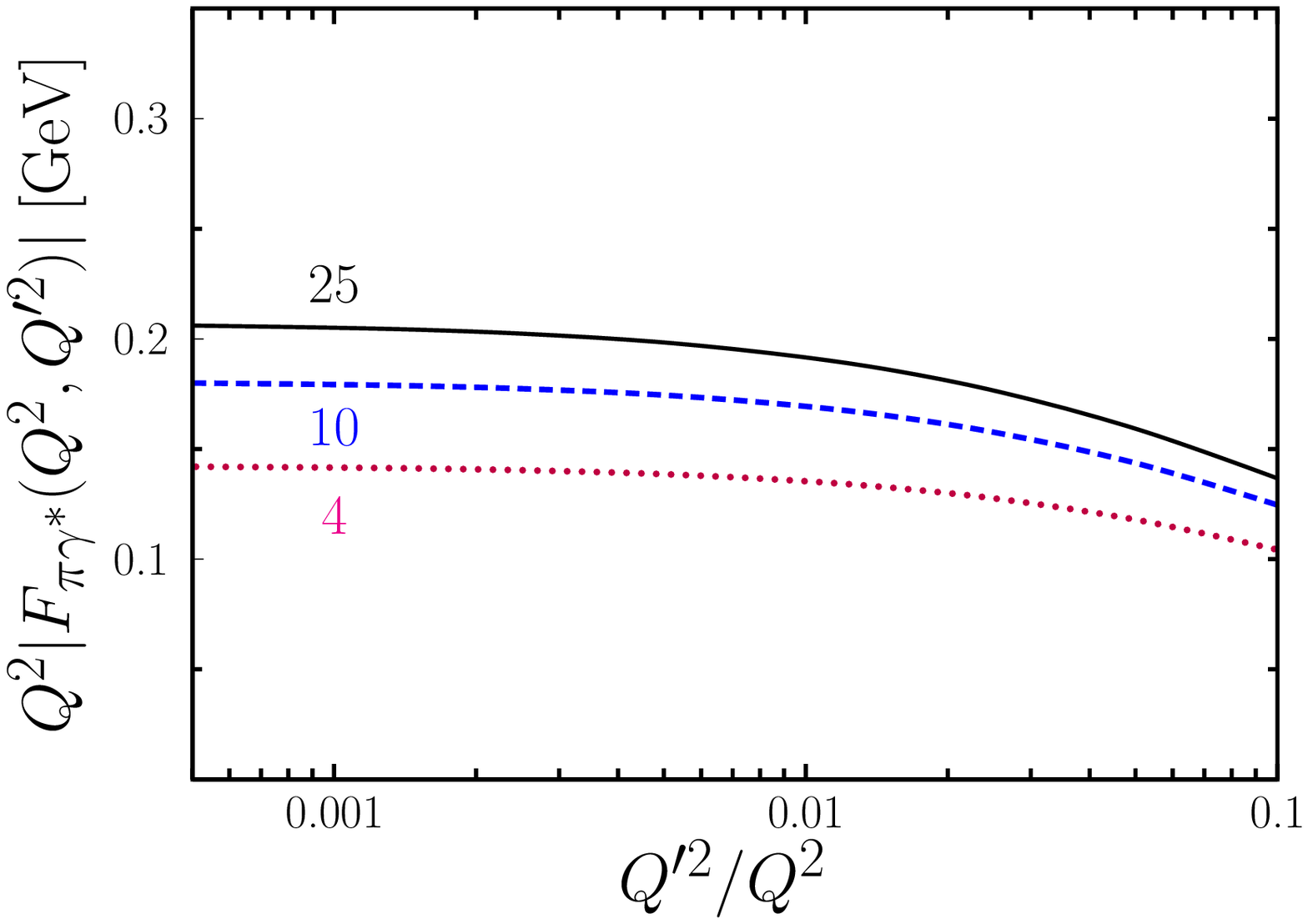} 
 \end{center}
\caption{The $\pi\gamma^*$ form factor versus the ratio of the two photon 
  virtualities for $Q^2=25\, (10, 4)\,\gev^2$ represented as solid (dashed,
  dotted) line. The parameters of the pion \wf{} are taken from fit 
  \req{eq:a4-fit}. (Colors online)}
\label{fig:gamma-star}
\end{figure}

The measurement of the space-like $\pi\gamma$ transition form factor is
performed in $e^+ e^-$ collisions with so-called single-tag events where
either the electron or the positron in the final state is detected. This 
method implies an integration over the spectrum of that exchanged photon 
which is emitted from the undetected lepton, up to a value of, say,
$Q^{\prime\, 2}$ set by an experimental cut. Actually for the BaBar 
experiment the cut is $Q^{\prime\, 2} < 0.18\,\gev^2$. Thus, strictly
speaking, one measures the form factor for the transition from a quasi-real 
photon with an effective virtuality $Q^{\prime\, 2}_{\rm eff}$ less than the 
cut value, to the pion. The question arises how good does this measured 
form factor approximate the one for the transition from a real photon to 
the pion. In order to examine this issue $F_{\pi\gamma^*}$ is evaluated for 
small values of the ratio $Q^{\prime\, 2}/Q^2$ using \req{eq:pi-gamma*} and 
the pion \wf{} parameters from fit \req{eq:a4-fit}. As one may see from Fig.\ 
\ref{fig:gamma-star} where the results for a sample set of $Q^2$ values 
are shown, the transition form factor depends on $Q^{\prime\, 2}$ mildly 
(at least within the MPA) as long as the ratio $Q^{\prime\, 2}/Q^2$ is smaller 
than about 0.01. At this value of the ratio of the two virtualities the form 
factor is reduced by about $5\%$ as compared to its value at the real photon 
limit while at a value of 0.1 of the ratio, the form factor is smaller by 
about $30\%$.  

\section{Remarks on the time-like transition form factors}
\label{sec:timelike}
In the context of the large $Q^2$ behavior of the transition form factors one
may also consider the recent measurement of the time-like $\eta\gamma$ and 
$\eta^\prime\gamma$ form factors at $s=112\,\gev^2$ by the BaBar collaboration 
\ci{BABAR06}: 
\be 
s|F_{\eta\gamma}|\=0.229\pm 0.031 \,\gev\,, \qquad 
       s|F_{\eta^\prime\gamma}|\=0.251\pm 0.021\, \gev\,.
\label{eq:BABAR-tl}
\ee
Comparing these data with the predictions for the space-like form factors
which are shown in Figs.\ \ref{fig:eta-etap} and \ref{fig:comp}, 
one notices that for the case of the $\eta$ the experimental time-like value
is much larger than the predicted space-like one which amounts to
0.169 at $Q^2=112\,\gev^2$. For the case of the $\eta^\prime$, on the
other hand, the experimental time-like value practically falls together 
with theoretical result of 0.258 for the space-like form factor. 

A detailed study of the pseudoscalar meson-photon transition form factors
in the time-like region is beyond the scope of the present article. Within 
the MPA however, one expects slightly different values for the time-like
and space-like transition form factors: The space-like propagator 
$[x Q^2+\vk^2]^{-1}$ is to be replaced by 
$[-x s+\vk^2 -\imath \epsilon]^{-1}$ in the time-like region or, in
$b$-space, $K_0(\sqrt{x} Qb)$ in \req{eq:th} by $i \pi/2
H_0^{(1)}(\sqrt{x} sb)$ with $H_0^{(1)}$ being the zeroth order Hankel
function. The pole of the propagator now occurs within the range of
integration and in general leads to an enhancement as well as a phase of the
form factor. For the case of the electromagnetic form factor of the pion this
phenomenon has been pointed out and studied in some detail by  Gousset and
Pire \ci{gousset}. The analytic continuation of the Sudakov factor from 
the space-like to the time-like region, which is necessary too, is not well 
understood. It probably leads to an oscillating phase \ci{gousset,magnea}.

In order to give an admittedly rough estimate of the expected size of the 
time-like form factor it is followed \ci{gousset} and the time-like propagator 
is combined with the space-like Sudakov factor. With this receipe one finds 
the absolute values 0.25 and $0.24\,\gev$ for the scaled time-like $\pi\gamma$ 
transition form factor at $s=3$ and $100\,\gev^2$, respectively. The
corresponding ratios of the time and space-like transition form factors are 
$\simeq 1.9$ and $\simeq 1.1$. For $s$ larger than about $5\,\gev^2$ the 
time-like form factor is dominantly real, its imaginary part contributes less 
than about $10\%$ to the absolute value. The accumulation profile of the
absolute value of the time-like form factor is displayed in Fig.\ 
\ref{fig:accumulation}. Although the pole of the propagator occurs in the 
end-point region (at $\xb=k_\perp/s$) the profile is only mildly softer than 
that of the space-like form factor, i.e.\ also the time-like form factor is 
mainly fed by contributions from regions where the renormalization scale is 
sufficiently large. In the light of this feature one may consider the results 
for the time-like transition form factor obtained within the MPA as
a tolerable estimate. 

Along these lines one can also compute the $\eta\gamma$- and
$\eta^\prime\gamma$ transition form factors in the time-like region. 
Using the parameters quoted in \req{eta-fit} one finds that at 
$s=3\,\gev^2$ the absolute values of the time-like form factors are about a 
factor of 2 larger than the space-like form factors. At $s=112 \,\gev^2$ 
the results are $s|F_{\eta\gamma}|\simeq 0.17\,\gev$  and  
$s|F_{\eta^\prime\gamma}|\simeq 0.28\,\gev$. While for the case of the
$\eta^\prime$ there is rough agreement with experiment \req{eq:BABAR-tl} within 
errors, is the $\eta\gamma$ form factor too small by about two standard
deviations.  

\section{Concluding remarks}
\label{sec:summary}
In this paper an analysis of the form factors for the transitions from
a photon to a pseudoscalar meson is presented. The analysis is performed 
within the MPA which bases on $\vk$ factorization. It is shown that due 
to the Sudakov suppressions which are an important ingredient of the MPA 
and which represents radiative corrections in next-to-leading-log 
approximation summed to all orders of perturbation theory, higher order 
Gegenbauer terms of the meson's \da{} are suppressed at low $Q^2$. 
In fact, the combined effect of the hard scattering kernel and the Sudakov
factor leads to a series of power suppressed terms which stem from the 
soft regions. The intrinsic $\vk$-dependence generates a second series
of power corrections which in contrast to the first series, do not grow
with the Gegenbauer index. The interplay of these two ingredients results 
in a remarkable feature of the MPA - the transition form factors are only 
affected by the few lowest Gegenbauer terms of the \da{}, the higher 
ones do practically not contribute. How many Gegenbauer terms are relevant 
depends on the range of $Q^2$ considered: In the $Q^2$ range covered by the 
CLEO data \ci{CLEO} ($<10\,\gev^2$) it suffices to use just the asymptotic 
\da{} in order to fit the CLEO data. With the BaBar data 
\ci{babar09,BaBar-prel} at disposal, covering the unprecedented large range 
$4\,\gev^2 < Q^2 < 35\,\gev^2$, the next or the next two Gegenbauer terms 
have to be taken into account or, turning the argument around, can be 
determined from an analysis of the data on the transition form factors. 
Indeed this is what has been done in this work. From the present analysis it 
turns out that for the case of the pion a fairly strong contribution from 
$a_2$ is required by the data while for the $\eta$ and $\eta^\prime$ much 
smaller deviations from the asymptotic \da{} are needed. For these cases the 
results from a previous calculation within the MPA \ci{feldmann97} are already 
in fair agreement with the BaBar data, nearly perfect for the $\eta$, slightly 
worse for the $\eta^\prime$. Comparing the $\pi\gamma$ form factor with the  
$\eta\gamma$ or more precisely the $\eta_8\gamma$ one, one observes a strong 
breaking of flavor symmetry in the ground-state octet of the pseudoscalar
mesons. The difference between the two form factors is larger than the 
respective decay constants. In other processes involving pseudoscalar mesons 
such large flavor symmetry violations have not been observed. With regard to 
the theoretical importance of the transition form factors, in particular the 
role of collinear factorization a remeasurment, e.g. by the BELLE
collaboration, would be highly welcome. 

One may wonder what the implications of the new \da s for the pseudoscalar
mesons are for other hard exclusive processes. A detailed investigation of
this issue is beyond the scope of the present paper. It has however been 
checked that for the pion's electromagentic form factor there is no 
substantial change. The perturbative contribution to it still amounts to 
only about a third of the experimental value of the form factor which is 
measured only at low $Q^2$ \ci{Fpi2}. The perturbative contribution is 
slightly increasing with growing $Q^2$ now and not as flat as in 
\ci{jakob93}. In any case for $Q^2$ less than $10\,\gev^2$ the
differences are marginal.

{\bf Acknowledgements}
It is a pleasure to thank Volodya Braun and Markus Diehl for their
interest in this work and numerous valuable comments. Discussions with 
Vladimir Druzhinin and Andreas Sch\"afer are also gratefully acknowledged.
This work is supported  in part by the BMBF, contract number 06RY258.\\
\begin{appendix}
\renewcommand{\theequation}{A.\arabic{equation}}
\section{Appendix}
\label{sec:appendix}
\setcounter{equation}{0}
In this appendix details of the Sudakov factor are presented. 
The Sudakov exponent $S$ which comprises the characteristic double logarithms
produced by overlapping collinear and soft divergencies (for massless quarks)
has been calculated in \ci{botts89}. In axial gauge, $n\cdot A=0$, these 
overlaps arise in general from two-particle reducible Feynman graphs where,
before the hard interaction, the gluon is exchanged between the quark and 
antiquark of the meson or emitted from and reabsorbed by either the quark or
the antiquark. The Sudakov factor can therefore be analyzed
indepedently of the physical process and can be viewed as part of the meson
\wf{}. For the case of interest, only the two-particle reducible graphs
occur anyway. As shown in \ci{botts89} for a quark-antiquark system the Sudakov
exponent reads~\footnote{
Musatov and Radyushkin \ci{musatov} claim that, due to kinematical properties
the Sudakov functions should be summed as $s({x},b,Q)+s(\sqrt{\xb},b,Q)$ where
it is assumed that the real photon is attached to the antiquark line. Doing so
the universality of the \wf{} (including the Sudakov factor) is broken.
This alternative possibility only leads to tiny numerical differences in the form
factor. The predictions change by less than $1\%$.}
\ba
S\= s(x,b,Q) + s(1-x,b,Q) + 2\,\int_{\muF}^{\muR}
\frac{d\bar{\mu}}{\bar{\mu}}\,\gamma_q(\als(\bar{\mu}))\,.
\label{eq:S}
\ea
The impact parameter $\vbs$, canonically conjugated to $\vk$, is the spacial
separation of quark and antiquark. The Sudakov function $s(\xi,b,Q)$ where 
$\xi$ is either $x$ or $1-x$, is given by 
\be
s(\xi,b,Q)\=\frac12\,\int_{C_1/b}^{C_2\xi Q}\,\frac{d\mu}{\mu}\,\left\{
     2 \ln{\Big(\frac{C_2\xi Q}{\mu}\Big)} {\cal A}(C_1,g(\mu))
   + {\cal B}(C_1,C_2,g(\mu)) \right\}\,.
\label{eq:s-integral}
\ee
An appropriate choice of the gauge vector $n$ has been made in order to obtain
this result. The function ${\cal B}$ has been calculated to order $\als$ in
the $\overline{\rm MS}$ scheme explicitly. The function ${\cal A}$ arises from 
the use of the renormalization group equation which is applied in order to 
absorb all the $Q^2$ dependence into the scale of the coupling constant. It 
has been calculated to order $\als^2$:
\be
{\cal A}\=\frac{\als}{\pi}\,A^{(1)} + \Big(\frac{\als}{\pi}\Big)^2\,A^{(2)} + {\cal
  O}(\als^3)
\ee
with
\be
A^{(1)}\=C_F\,, \qquad 
 A^{(2)} \= \frac{C_F}2\left[\frac{67}{6}-\frac{\pi^2}2 -\frac59 n_f 
+ \beta_0 \ln{(C_1 e^{\gamma_E}/2)}\right]\,,
\ee              
where $\gamma_E$ is the Euler constant. The constants $C_1$ and $C_2$ are free
parameters which may be chosen in such a way that large logarithms from higher
orders in the perturbative expansion of ${\cal B}$ are avoided. The
conventional choice is
\be
C_1\=1\,, \qquad C_2\=\sqrt{2}\,.
\label{eq:choice}
\ee  
Other choices of the $C_i$ have been discussed in \ci{kim98}. 

The Sudakov function has been explicitly given in \ci{botts89} first.   
Later repetitions of this calculation \ci{bolz95,descotes02} led
to a slightly different result which is quoted here 
\ba
  s(\xi,b,Q)
&=& \frac{2 C_F}{\beta_{0}}\;\left[
   \qh \ln{\Big(\frac{\qh}{\bh}\Big)} -\qh + \bh \right] \nn\\
 &+&  \frac{C_F \beta_{1}}{\beta_0^3}\,\left[
   \qh\Big(\frac{\ln{(2\qh)} + 1}{\qh}
           -\frac{\ln{(2\bh)} + 1}{\bh}\Big) + \frac12\ln^2(2\qh) - \frac12\ln^2(2\bh)
                \right]  \nn \\
 &+&  4\frac{A^{(2)}}{\beta_0^2}\,
           \left[\frac{\qh -\bh}{\bh} -\ln{\Big(\frac{\qh}{\bh}\Big)}\right] 
+ \frac{C_F}{\beta_0}\ln{\Big(\frac{C_1^2 e^{2\gamma_E
      -1}}{C_2^2}\Big)}\,\ln{\Big(\frac{\qh}{\bh}\Big)}\,.
\label{eq:s}
\ea
The variables $\qh$ and $\bh$ are defined by
\be
\qh\= \ln{\Big(\frac{C_2\xi Q}{{2}\LQCD}\Big)}\,, \qquad 
\bh\=\ln{\Big(\frac{C_1}{b\LQCD}\Big)}\,.
\label{eq:def-qh-bh}
\ee
The last term in \req{eq:s} represents the integrated function ${\cal B}$. 
The differences between the various results for $s$ used earlier 
\ci{botts89,li92,jakob93} and \req{eq:s} are numerically small, of the order
of a few percent. It should be noted that the last term in \req{eq:s} is twice
as large as in \ci{kim98}. Also this discrepancy has little effect on the form
factor.
  
The integral in \req{eq:S} arises from the application of the renormalization 
group equation. It combines the effects of the renormalization
group equation on the \wf{} (with the factorization scale $\muF$)
and on the hard scattering amplitude involving the renormalization scale
$\muR$. The evolution from one scale to another is controlled
by the anomalous dimension of the quark's \wf{} in axial gauge 
\ci{sop80} $\gamma _{q}=-\als/\pi\;+\;{\cal O}(\als^2)$. It leads to  
\be
\int_{\muF}^{\muR}
\frac{d\bar{\mu}}{\bar{\mu}}\,\gamma_q(\als(\bar{\mu}))\=
 \frac2{\beta_0}\,\ln\frac{\ln{\big(\muF^2/\LQCD^2\big)}}{\ln{\big(\muR^2/\LQCD^2\big)}}
\label{eq:RGN}
\ee
with the help of the 1-loop result for $\als$. Its use is consistent with the
derivation of \req{eq:s}. In particular the terms in the first line of 
\req{eq:s} are calculated from the 1-loop $\als$ and these terms dominate the
Sudakov function. 
\end{appendix}


\end{document}